\documentclass[ba,review]{imsart}
 
\usepackage{amsfonts,amsmath,acronym,amssymb,theorem,graphicx,hyperref,color,natbib}
\usepackage{bm}
\usepackage{mathtools}
\usepackage{booktabs}
\usepackage{mathrsfs} 
\usepackage{mathptmx} 
\usepackage{multirow}
\usepackage{color}
\usepackage{mathrsfs}
\usepackage{caption}
\usepackage{subcaption}
\usepackage{latexsym}
\usepackage{nicefrac}
\usepackage{verbatim}
\usepackage{lineno}
\usepackage[ruled]{algorithm2e}

\newtheorem{exemp}{Example}[section]

\begin{document}

\begin{frontmatter}

\title{Bayesian Estimation Approach for Linear Regression Models with Linear Inequality Restrictions\protect\thanksref{T1}}
\title{}
\runtitle{}
\thankstext{T1}{Solmaz  Seifollahi, Department of Statistics, Faculty of Mathematical Science, University of Tabriz, Tabriz, Iran, {\sf s.seifollahi@tabrizu.ac.ir}, Kaniav Kamary, CentraleSup\'elec, Universit\'e Paris-Saclay, Gif-sur-Yvette, France, {\sf kaniav.kamary@centralesupelec.fr}, Hossein Bevrani, Department of Statistics, Faculty of Mathematical Science, University of Tabriz, Tabriz, Iran, {\sf bevrani@gmail.com}.
}
 
\begin{aug}
 \author{\snm{{\sc Solmaz  Seifollahi,}}}
 \affiliation{University of Tabriz, Tabriz, Iran}
 \author{\snm{{\sc Kaniav Kamary,}}}
 \affiliation{CentraleSup\'elec, Universit\'e Paris-Saclay, France}
 \author{\snm{{\sc Hossein Bevrani}}}
 \affiliation{University of Tabriz, Tabriz, Iran}
\end{aug}

\begin{abstract} 
Univariate and multivariate general linear regression models, subject to linear inequality constraints, arise in many scientific applications. The linear inequality restrictions on model parameters are often available from phenomenological knowledge and motivated by machine learning applications of high-consequence engineering systems \citep{Agrell,Veiga}.
Some studies on the multiple linear models consider known linear combinations of the regression coefficient parameters restricted between upper and lower bounds. In the present paper, we consider both univariate and multivariate general linear models subjected to this kind of linear restrictions.
So far, research on univariate cases based on Bayesian methods is all under the condition that the coefficient matrix of the linear restrictions is a square matrix of full rank. This condition is not, however, always feasible. Another difficulty arises at the estimation step by implementing the Gibbs algorithm, which exhibits, in most cases, slow convergence. 
This paper presents a Bayesian method to estimate the regression parameters when the matrix of the constraints providing the set of linear inequality restrictions undergoes no condition. For the multivariate case, our Bayesian method estimates the regression parameters when the number of the constrains is less than the number of the regression coefficients in each multiple linear models.
We examine the efficiency of our Bayesian method through simulation studies for both univariate and multivariate regressions. After that, we illustrate that the convergence of our algorithm is relatively faster than the previous methods. Finally, we use our approach to analyze two real datasets.
\end{abstract}


\end{frontmatter} 
 
\section{Introduction}
Recently, the univariate and multivariate general linear regression models (MGLM) have found a wide range of applications in various fields.
\cite{izenman} gives an example from machine learning theory, which includes discrimination and classification problems as well as artificial intelligence. Another one can be found in psychology and education \citep{timm}, as well as in discovering gene expression patterns \citep{zapala}.
 We can also find the application of MGLMs in econometrics, genomics \citep{kim2,lee,wang}, chemometrics \citep{srivastava}, bioinformatics \citep{meng}, and other fields.
 
Some recent studies focus on the possibility of having uncertain prior information on some of the model parameters. They often incorporate the uncertainty in the model by assuming that the regression coefficient vector ($\pmb{\beta}$) is subject to a set of linear inequality constraints. For example, they suppose $H\pmb{\beta}\leq \pmb{G}$ where $H$ is called the restriction matrix, or in a more general case, they assume that some coefficient parameters or known linear combination of the coefficient parameters are restricted between upper and lower bounds. In applied econometrics, we deal with the situation in which some coefficient parameters should be non-negative or non-positive \citep{Pindyck,Bails}.  
In hyperspectral imaging \citep{Manolakis}, due to physical considerations, the coefficient parameters should be non-negative. \citet{Zhu1} provides another example of these restrictions in geodesy.

Statistical inference of the linear regression model is commonly carried out using the ordinary least square estimator, which coincides with the maximum likelihood estimator.  
However, the traditional least square method may not always satisfy the restrictions and so various studies have been done to improve the MLEs while linear equality restrictions are held (see for instance \cite{Ahmed,Bahadir,kim1} and \cite{chitsaz1,chitsaz2}).
In the classical inference of the univariate regression models, \citet{Judge} gave the inequality constrained least square (ICLS) estimate of the regression coefficients using the Dantzig-Cottle algorithm. They showed that for a sufficiently large sample, this estimator mimics the ordinary least square estimator. Then \cite{escobar2}, and \cite{ohtani} obtained some properties of this estimator such as the bias \citep{escobar1}, the mean square error, and the efficiency over inequality restrictions. 

When the preliminary information represented by the restrictions is uncertain, investigators often implement a pre-test to examine the validity of the restrictions. A great deal of research has been done on testing under inequality restrictions for linear models such as \citet{Wolak1, Wolak2}, \citet{Gourieroux}, \citet{Geng}, \citet{Fonseca} and \citet{ZHU2}. This paper focuses on Bayesian methods whose estimation procedure uses the information obtained from observations and historical data or other sources. \citet{Geweke1} had a Bayesian approach to the problem of the univariate linear model subject to the inequality restrictions on the coefficient parameters. 
He considered a prior distribution on the parameters consisting of a non-informative distribution and an indicator function presenting the restrictions on parameters. \citet{Geweke1} used the importance sampling method to estimate the parameters. However, \citet{Geweke1}'s process of estimating might be slow in terms of the MCMC convergence when the number of the coefficient parameters increases or when the posterior probability of the inequality restrictions is low. 
Furthermore, \cite{Geweke3} studied the linear inequality restrictions and expanded the study of \citet{Davis}, \cite{Chamberlain} and \citet{Leamer} by implementing the Gibbs algorithm to estimate the coefficient parameters. This method converges faster and gives more accurate approximations to the posterior moments.
However, \cite{Geweke3}'s algorithm is practicable whenever the restriction matrix is a square and invertible matrix. This condition is possible when the number of linear restrictions is equal to the dimension of the coefficient vector. The invertibility of the restriction matrix is also possible when the summation of the number of the linear constrains is equal to the coefficient vector length. In other cases, we fail to use the \cite{Geweke3}'s algorithm. 
\citet{Rodriguez1,Rodriguez2}'s also used a Gibbs sampler algorithm that needs to transform the general restrictions into a one-sided inequality.
This paper introduces a Bayesian method of estimating the univariate and multivariate regression parameters when the restriction matrix has any form (a square matrix or non-square matrix). For the multivariate regression model, the number of restrictions on coefficient parameters of each multiple linear models is less than the number of coefficient parameters. 
By partitioning the restriction matrix into two matrixes, one of which is full rank and non-singular, we reduce the number of the parameters estimated by the Gibbs algorithm. Due to this strategy, our algorithm is less time-consuming compared to the previously presented algorithms.

The structure of the remainder of the text is as follows: Section \ref{models} covers a description of our univariate and multivariate regression models. In Section \ref{section:model}, we then proceed with the method of the partitioning of the restriction matrix, the choice of the prior distributions for the model parameters, and the posterior distributions. After that, we describe our MCMC algorithm. In Section \ref{simulation}, we illustrate the performance of our method through simulation studies for both univariate and multivariate regressions. Then we perform a comparison between the performance of our approach and Geweke's method for the univariate case. Finally, we provide the analysis of two real datasets in Section \ref{real.study}.

\section{Linear Regression Models}\label{models}
\subsection{Univariate case : }
Univariate linear regression model is defined as follows;
\begin{equation}
y_i=X_i\pmb{\beta}+\epsilon_i, \qquad i:1, 2, \cdots, n
\end{equation}
where $\pmb{\beta} \in \mathbb{R}^p$ is the coefficient vector and the error term $\epsilon_i$ has a normal distribution, $\epsilon_i \sim \mathcal{N}(0,\sigma^2)$. 
This model in matrix form is then given by:
\begin{equation}\label{m1}
Y=\pmb{X}\pmb{\beta}+\epsilon,
\end{equation}
where $Y=(y_1, y_2, \cdots, y_n)' $ is the response vector, $ \pmb{X}$ is the design matrix of size $ n\times p $, $\pmb{\beta}= (\beta_1, \beta_2, \cdots, \beta_p)'$ is a $p-$length vector of the regression coefficients and $\epsilon=(\epsilon_1, \epsilon_2, \cdots, \epsilon_n)' $ is the error vector. In this paper, we consider a general framework in which the coefficient parameter vector $ \pmb{\beta} $ is subjected to a set of $ q$ independent linear inequality restrictions as 
\begin{equation}\label{rest}
 K \leq H \pmb{\beta} \leq G
\end{equation}
where $ H $ is a matrix of size $ q \times p $ with $ (q\leq p) $ and $ G $ and $ K $ are two vectors of length $ q $.

\subsection{Multivariate case :} 
Multivariate general linear models (MGLMs) are a generalized form of multiple linear models in which a set of explanatory variables or covariates is used to predict several response variables :
\begin{equation}\label{m2}
 \pmb{Y}_{(n, k)}= \pmb{X}_{(n, p)}\pmb{B}_{(p, k)}+ \pmb{E}_{(n, k)}
\end{equation}
where $\pmb{Y}$ is the $n\times k$ response matrix; $\pmb{X}$ is a fixed and know matrix of size $n\times p$ such that all entries of the first column are $1$s and it contains $k-1$ predictors; $\pmb{B}$ is a $p\times k$ matrix of the regression parameters (one column for each response variable); and $\pmb{E}$ is a matrix of the model errors. We assume that $\pmb{X}$ and $\pmb{E}$ are independent. 
If $\epsilon_i'$ represents the $i$th row of the error matrix $\pmb{E}$, then we suppose that 
$$\epsilon_i' \sim \mathcal{N}_{k}(\pmb{0}, \pmb{\Sigma}),\quad \text{and} \quad \forall i\neq j : \epsilon_i'~ \text{and}~ \epsilon_j'~ \text{are independent}.$$ 
where $\pmb{\Sigma}$ is a $k \times k$ non-singular covariance matrix. We can then write
\begin{equation}
 vec(\pmb{E})= \mathcal{N}(\pmb{0}, \pmb{\Sigma}\otimes \pmb{I}_n),
\end{equation}
in which $\pmb{0}$ is a vector of length $nk$ of $0$s, the $vec$ ravels the error matrix column-wise into a vector and $\pmb{I}_n$ is the identity matrix, and $\otimes$ stands for the right knocker product operation. The general form of the restrictions can be considered as follows in this case:
\begin{equation*}
 \pmb{K}\leq \pmb{R}\pmb{B}\leq \pmb{G}
\end{equation*}
where $\pmb{R},  \pmb{K}$ and $\pmb{G}$ are the matrixes of sizes $q\times p$, $q\times k$ and $q\times k$, respectively.

\section{Bayesian Inference of the Models}\label{section:model}
\subsection{Prior Specification}
When making a Bayesian inference, the prior distribution should take into account all prior information that we have on the model parameter. If the elements of the matrix $H$,  the vector $\pmb{\beta}$, $G$  and $K$ are defined as

\begin{align*}
H_{(q,p)} = 
\begin{pmatrix}
h_{1,1} & h_{1,2} & \cdots & h_{1,p} \\
h_{2,1} & h_{2,2} & \cdots & h_{2,p} \\
\vdots  & \vdots  & \ddots & \vdots  \\
h_{q,1} & h_{q,2} & \cdots & h_{q,p} 
\end{pmatrix}, \quad \pmb{\beta}= \begin{pmatrix}
\beta_{1} \\
\beta_{2} \\
\vdots\\
\beta_{p} 
\end{pmatrix} , \quad G_{(q,1)} = 
\begin{pmatrix}
g_{1} \\
g_{2} \\
\vdots\\
g_{q} 
\end{pmatrix}, \quad K_{(q,1)} = 
\begin{pmatrix}
k_{1} \\
k_{2} \\
\vdots\\
k_{q}
\end{pmatrix}
\end{align*}
\\
the equation $K \leq H\pmb{\beta}\leq G$ can then represent a system of $q$ linear equations with $p$ unknowns as
\begin{align}\label{eq:system}
k_{1} \leq h_{1,1}\beta_1 + h_{1,2}\beta_2 + \cdots + h_{1,p}\beta_p& \leq g_{1} \nonumber\\
k_{2} \leq h_{2,1}\beta_1 + h_{2,2}\beta_2 + \cdots + h_{2,p}\beta_p& \leq g_{2}\nonumber \\
\vdots \nonumber \\
k_{q} \leq h_{q,1}\beta_1 + h_{q,2}\beta_2 + \cdots + h_{q,p}\beta_p& \leq g_{q}
\end{align} 
in which for $i=1, \ldots, q$ and $j=1, \ldots, p$, the constants $h_{i,j}$ are the coefficients of the system, and $g_i$ and $k_i$ are the constant terms.

For the multivariate general linear models, the restriction system and the matrices $\pmb{R}$, $\pmb{B}$, $\pmb{K}$  and $\pmb{G}$ are defined as following :
\begin{align*}
\pmb{R}_{(q, p)} =
\begin{pmatrix}
R_{11} & R_{12} & \cdots & R_{1p} \\
R_{21} & R_{22} & \cdots & R_{2p} \\
\vdots  & \vdots  & \ddots & \vdots  \\
R_{q1} & R_{q2} & \cdots & R_{qp}
\end{pmatrix}, \quad
\pmb{B}_{(p, k)}= \begin{pmatrix}
\beta_{11} & \beta_{12} & \cdots & \beta_{1k} \\
\beta_{21} & \beta_{22} & \cdots & \beta_{2k} \\
\vdots  & \vdots  & \ddots & \vdots  \\
\beta_{p1} & \beta_{p2} & \cdots & \beta_{pk}
\end{pmatrix}
\end{align*}
and
\begin{align*}
\pmb{G}_{(q, k)} =
\begin{pmatrix}
G_{11} & G_{12} & \cdots & G_{1k} \\
G_{21} & G_{22} & \cdots & G_{2k} \\
\vdots  & \vdots  & \ddots & \vdots  \\
G_{q1} & G_{q2} & \cdots & G_{qk}
\end{pmatrix}, \quad 
\pmb{K}_{(q, k)} =
\begin{pmatrix}
K_{11} & K_{12} & \cdots & K_{1k} \\
K_{21} & K_{22} & \cdots & K_{2k} \\
\vdots  & \vdots  & \ddots & \vdots  \\
K_{q1} & K_{q2} & \cdots & K_{qk}
\end{pmatrix}
\end{align*}
Note that in this case, the equation $ \pmb{K}\leq \pmb{R}\pmb{B}\leq \pmb{G}$ can represent a system of $q \times k (q\leq p) $ linear equations with $p \times k$ unknown parameters.

For the model \eqref{m1}, in order to determine the prior distribution of $\pmb{\beta}$ under the condition \eqref{eq:system}, we first partition the matrix $H$ into two sub-matrixes $H_S$ and $H_{S'}$ by a collection colgroups. Suppose that $P=\{1, \ldots,p \}= S\cup S'$ is a set of the indices of the matrix $H$ and  $S=\{j : h_{i,j} \in H_S;  i=1, \ldots, q \}\subset P$ is a subset of the indices of the columns of $H$ for which the block $H_S$ is a full rank matrix. The matrix $H_S$ is then a $q\times q$ full rank block of $H$, and $H_{S'}$ is a $q\times (p-q)$ sub-matrix including the rest of $H$. The cardinality of $S$ is then $|S|=q$ and that of $S'$ is $|S'|=p-q$. In the same way, the coefficient vector $\pmb{\beta}$ is partitioned into two sub-vectors, a $q-$length vector $\pmb{\beta}_S$ and a $(p-q$)-length vector $\pmb{\beta}_{S'}$.
We can therefore rewrite the system \eqref{eq:system} as follows
\begin{align}\label{ss}
K \leq
\begin{pmatrix}
{H_S}_{(q,q)} & {H_{S'}}_{(q,p-q)}
\end{pmatrix} \begin{pmatrix}
{\pmb{\beta}_S}_{(q,1)} \\
 {\pmb{\beta}_{S'}}_{(p-q,1)}
\end{pmatrix}&\leq G\nonumber\\
K \leq H_S\pmb{\beta}_S + H_{S'}\pmb{\beta}_{S'} \leq G, \quad \text{and then }~K - H_{S'}\pmb{\beta}_{S'} \leq H_S\pmb{\beta}_S& \leq G - H_{S'}\pmb{\beta}_{S'}
\end{align} 

Under the definition \eqref{ss}, the prior distribution of the parameters $\pmb{\beta}_S$ depends on that of $\pmb{\beta}_{S'}$. In other words, if $\pmb{\beta}_{S'}\in \mathbb{R}^{p-q}$, then the support of $\pmb{\beta}_S$ is a subspace of $\mathbb{R}^q$ for which 
 $K - H_{S'}\pmb{\beta}_{S'} \leq H_S\pmb{\beta}_S \leq G - H_{S'}\pmb{\beta}_{S'}$ holds.
 
In the same manner, we can then generalize the partitioning procedure of the restriction matrix $\pmb{R}$ and the parameter matrix $\pmb{B}$ for the multivariate model as follows : 
 \begin{equation*}
\pmb{K} \leq
  \begin{pmatrix}
  {\pmb{R}_S}_{(q,q)}, & {\pmb{R}_{S'}}_{(q,(p-q))}
\end{pmatrix} \begin{pmatrix}
{\pmb{B}_S}_{(q, k)} \\
 {\pmb{B}_{S'}}_{((p-q), k)}
\end{pmatrix}\leq \pmb{G}
\end{equation*}
\begin{equation}\label{eq:system:matrix}
 \pmb{K}\leq \pmb{R}_S\pmb{B}_S + \pmb{R}_{S'}\pmb{B}_{S'} \leq \pmb{G}, \quad \text{and then }~ \pmb{K} - \pmb{R}_{S'}\pmb{B}_{S'} \leq \pmb{R}_S\pmb{B}_S  \leq \pmb{G} - \pmb{R}_{S'}\pmb{B}_{S'}
\end{equation}
 
 A classical choice for the prior distribution of the regression coefficients $\pmb{\beta}_{S'}$ and $\sigma^2$ is the conjugate prior distributions given by
\begin{align}\label{prior1}
\sigma^2 & \sim \mathcal{IG}(\dfrac{a}{2},\dfrac{b}{2}),\nonumber\\
 \pmb{\beta}_{S'}| \sigma^2 & \sim \mathcal{N}(\mu_{\pmb{\beta}_{S'}}, \sigma^2C_{\pmb{\beta}_{S'}}),
\end{align}
where $a, b$ and $\mu_{\pmb{\beta}_{S'}}$ are supposed to be known and $C_{\pmb{\beta}_{S'}}$ is a $(p-q)\times (p-q)$ positive definite symmetric and known matrix. 
In order to choose a prior distribution for the parameter $\pmb{\beta}_S$ that depends on the prior information in \eqref{ss}, we consider 
\begin{equation}\label{prior2}
\pmb{\beta}_{S}| \pmb{\beta}_{S'},\sigma^2 \sim \mathcal{N}_{\mathcal{T}}(\mu_{\pmb{\beta}_{S}}, \sigma^2C_{\pmb{\beta}_{S}}), 
\end{equation}
in which
\begin{equation}\label{support}
 \mathcal{T}= \{ \pmb{\beta}_{S} \in \mathbb{R}^{q} | K - H_{S'}\pmb{\beta}_{S'} \leq H_{S} \pmb{\beta}_S \leq G - H_{S'}\pmb{\beta}_{S'}\big \}.
\end{equation}
$\mu_{\pmb{\beta}_{S}}$ is supposed to be known and $C_{\pmb{\beta}_{S}}$ is a $q\times q$ positive definite symmetric and known matrix. The distribution $\mathcal{N}_{\mathcal{T}}$ indicates a multivariate truncated normal distribution over $ \mathcal{T} $ (See  \citet{Horrace} for more details).
For the MGLM case, we consider an Inverse Wishart distribution for the matrix $\pmb{\Sigma}$ and a multivariate Gaussian distribution for the regression coefficients $\pmb{B}_{S'}$ and  $\pmb{B}_S$ as follows :

\begin{align}\label{prior21}
\pmb{\Sigma} & \sim \mathcal{IW}_k(r,\pmb{Q}),\nonumber\\
 \pmb{B}_{S'}| \pmb{\Sigma} & \sim \mathcal{N}(\pmb{M}_{\pmb{B}_{S'}}, \pmb{\Sigma}\otimes \pmb{D}_{\pmb{B}_{S'}}),
\end{align}
where $\pmb{M}_{\pmb{B}_{S'}}$ and $\pmb{D}_{\pmb{B}_{S'}}$ are a $(p-q)\times k$ matrix and a $(p-q)\times (p-q)$ positive definite symmetric matrix, respectively. Note that, both matrixes $\pmb{M}_{\pmb{B}_{S'}}$ and $\pmb{D}_{\pmb{B}_{S'}}$, are supposed to be known. We have then 
\begin{equation}\label{prior22}
\pmb{B}_{S}| \pmb{B}_{S'},\pmb{\Sigma} \sim \mathcal{N}_{\mathcal{T}}(\pmb{M}_{\pmb{B}_S}, \pmb{\Sigma}\otimes \pmb{D}_{\pmb{B}_S}),
\end{equation}
in which
\begin{equation}\label{support2}
 \mathcal{T}= \{ \pmb{B}_{S} \in \mathbb{R}^{q \times k} | {\pmb{K} - \pmb{R}_{S'}\pmb{B}_{S'} \leq \pmb{R}_S\pmb{B}_S \leq \pmb{G} - \pmb{R}_{S'}\pmb{B}_{S'}}\big \},
\end{equation}
and $\pmb{M}_{\pmb{B}_{S}}$ and $\pmb{D}_{\pmb{B}_S}$ are a $q\times k$ matrix and a $q\times q$ positive definite symmetric matrix, respectively, and supposed to be known.

\subsection{Posterior Distributions}
\subsubsection{Univariate linear regression model :} 
If we suppose that $\pmb{\beta}_{S;{S'}}$ is defined as 
$\pmb{\beta}_{S;{S'}}=\begin{pmatrix}
\pmb{\beta}_S \\ 
\pmb{\beta}_{S'}
\end{pmatrix},$ 
then $ \pmb{\beta}_{S;S'} $ is possibly equal to $ \pmb{\beta} $ or it may be a vector of the regression coefficients, $ \beta_j; j=1, \ldots, p $, whose elements are a permutation of the elements of $ \pmb{\beta} $. We therefore rewrite the model  \eqref{m1} according to $ \pmb{\beta}_{S;S'} $ in order to calculate the posterior distribution of the regression coefficients based on the prior distribution defined in \eqref{prior1} and \eqref{prior2}. Suppose that the columns of the explanatory variable matrix $ \pmb{X} $ are permuted according to the order of the indices of the elements of $ \pmb{\beta}_{S;S'}$ and
$\pmb{X}_{S;S'}=\left(\pmb{X}_{S(n,q)}\quad \pmb{X}_{S'(n,p-q)}\right)$ 
indicates this column-permuted matrix. Then the model  \eqref{m1} is therefore rewritten as
\begin{align}\label{m12}
Y&=\pmb{X}_{S}\pmb{\beta}_{S}+\pmb{X}_{S'}\pmb{\beta}_{S'} +\pmb{\epsilon} 
\end{align}
By using the prior distributions \eqref{prior1} and \eqref{prior2}, we obtain the posterior distributions for the parameters of the model \eqref{m12} as follows :
\begin{enumerate}
\item[-] The marginal posterior distribution of $\sigma^2$ : 
\begin{align}\label{cond1}
\sigma^2 | Y, \pmb{X}_{S;S'}&\sim \mathcal{IG}(\tilde{\nu}, \tilde{\eta})\\
\text{with}~~\tilde{\nu}&=\frac{n+a}{2} \nonumber\\
\text{and}~~\tilde{\eta}&=\frac{1}{2}\left[(b+ Y^T Y + \mu_{\pmb{\beta}_{S'}}^{T}C_{\pmb{\beta}_{S'}}^{-1} \mu_{\pmb{\beta}_{S'}} +\mu_{\pmb{\beta}_{S}}^{T}C_{\pmb{\beta}_{S}}^{-1} \mu_{\pmb{\beta}_{S}})-(\tilde{\mu}_{\pmb{\beta}_{S'}}^T\tilde{C}_{\pmb{\beta}_{S'}}^{-1}\tilde{\mu}_{\pmb{\beta}_{S'}})-W^T \tilde{C}_{\pmb{\beta}_S}W\right]\nonumber
\end{align}
where 
\begin{align*}
\tilde{C}_{\pmb{\beta}_{S'}}^{-1}&=\pmb{X}_{S'}^{T} \pmb{X}_{S'}+ C_{\pmb{\beta}_{S'}}^{-1}-(\pmb{X}_S^T\pmb{X}_{S'})^T \tilde{C}_{\pmb{\beta}_S}(\pmb{X}_S^T\pmb{X}_{S'})\nonumber\\
 \tilde{C}_{\pmb{\beta}_S}^{-1}&= (\pmb{X}_{S}^{T}\pmb{X}_{S}+ C_{\pmb{\beta}_{S}}^{-1} ) \nonumber\\
W&= \pmb{X}_{S}^{T} Y +  C_{\pmb{\beta}_{S}}^{-1}  \mu_{\pmb{\beta}_{S}} \nonumber\\
\tilde{\mu}_{\pmb{\beta}_{S'}}&=\tilde{C}_{\pmb{\beta}_{S'}}(C_{\pmb{\beta}_{S'}}^{-1}\mu_{\pmb{\beta}_{S'}} + \pmb{X}_{S'}^{T} Y-(\pmb{X}_{S'}^T\pmb{X}_S) \tilde{C}_{\pmb{\beta}_S}W)\nonumber
\end{align*}
\item[-] The conditional posterior distribution of $\pmb{\beta}_{S'}$: 
\begin{align}\label{cond2}
\pmb{\beta}_{S'}|\sigma ^2, Y, \pmb{X}_{S;S'}& \sim \mathcal{N}(\tilde{\mu}_{\pmb{\beta}_{S'}}, \sigma^2\tilde{C}_{\pmb{\beta}_{S'}})
\end{align}
\item[-] The conditional posterior distribution of $\pmb{\beta}_{S}$ which is given by
\begin{align}\label{cond3}
\pmb{\beta}_{S}|\pmb{\beta}_{S'}, \sigma^2, &Y, \pmb{X}_{S;S'} \sim \mathcal{N}_{\mathcal{T}}(\tilde{\mu}_{\pmb{\beta}_{S}}, \sigma^2\tilde{C}_{\pmb{\beta}_{S}})\\
\tilde{\mu}_{\pmb{\beta}_S}&= \tilde{C}_{\pmb{\beta}_S}(W-\pmb{X}_S^T\pmb{X}_{S'}\pmb{\beta}_{S'}).
\end{align}
\end{enumerate}
See Appendix \ref{Appendix1} for more details about the computation of the conditional posteriors obtained in \eqref{cond1}, \eqref{cond2} and \eqref{cond3}.
We call this Bayesian approach, \textbf{BKS-approach}. \\

\subsubsection{Multivariate general linear regression model :} In the same way as for the univariate case, if we assume that 
$\pmb{B}_{S;{S'}}=\begin{pmatrix}
\pmb{B}_S \\
\pmb{B}_{S'}
\end{pmatrix}$ and $\pmb{X}_{S;S'}=\left(\pmb{X}_{S(n,q)}\quad \pmb{X}_{S'(n,p-q)}\right)$, then the model \eqref{m2} can be rewritten as
\begin{align}\label{m22}
\pmb{Y}&=\pmb{X}_{S}\pmb{B}_{S}+\pmb{X}_{S'}\pmb{B}_{S'} +\pmb{E}.
\end{align}
and by using the prior distributions \eqref{prior21}, \eqref{prior22},we obtain the marginal distribution of $\pmb{\Sigma}$, the conditional posterior distributions of $\pmb{B}_{S'}$ and $\pmb{B}_{S}$ as follows:

\begin{equation}\label{post_sigma}
\pmb{\Sigma}|\pmb{Y}, \pmb{X}_{S;S'}\sim \mathcal{IW}_k (n+r,\pmb{\mathbb{V}})
\end{equation}
where
\begin{align*}
\pmb{\mathbb{V}} = \pmb{Q}+ \pmb{Y}^T \pmb{Y} + \pmb{M}_{\pmb{B}_{S'}}^{T} &\pmb{D}_{\pmb{B}_{S'}}^{-1} \pmb{M}_{\pmb{B}_{S'}}+ \pmb{M}_{\pmb{B}_S}^{T} \pmb{D}_{\pmb{B}_S}^{-1} \pmb{M}_{\pmb{B}_S}- \pmb{W}^T \tilde{\pmb{D}}_{\pmb{B}_S}\pmb{W}- \tilde{\pmb{M}}_{\pmb{B}_{S'}}^{T} \tilde{\pmb{D}}_{\pmb{B}_{S'}}^{-1} \tilde{\pmb{M}}_{\pmb{B}_{S'}}\\
\tilde{\pmb{D}}_{\pmb{B}_{S'}}^{-1}& = \pmb{D}_{\pmb{B}_{S'}}^{-1}+\pmb{X}_{S'}^T\pmb{X}_{S'}-(\pmb{X}_{S}^T\pmb{X}_{S'})^T \tilde{\pmb{D}}_{\pmb{B}_{S}}(\pmb{X}_{S}^T\pmb{X}_{S'}) \\
\tilde{\pmb{D}}_{\pmb{B}_{S}}^{-1}& = \pmb{D}_{\pmb{B}_S}^{-1} + \pmb{X}_{S}^T \pmb{X}_{S}\nonumber\\
\tilde{\pmb{M}}_{\pmb{B}_{S'}}& = \tilde{\pmb{D}}_{\pmb{B}_{S'}}\left(\pmb{X}_{S'}^T \pmb{Y}+ \pmb{X}_{S'}^T\pmb{M}_{\pmb{B}_{S'}}-(\pmb{X}_{S}^T\pmb{X}_{S'})\tilde{\pmb{D}}_{\pmb{B}_{S}} \pmb{W}\right)\\
\pmb{W}& = \pmb{X}_{S}^T \pmb{Y}+ \pmb{X}_{S}^T \pmb{M}_{\pmb{B}_S}
\end{align*}
and
\begin{equation}\label{post_Bsp}
\pmb{B}_{S'}|\pmb{\Sigma},\pmb{Y}, \pmb{X}_{S;S'} \sim \mathcal{N}(\tilde{\pmb{M}}_{\pmb{B}_{S'}}, \pmb{\Sigma}\otimes \tilde{\pmb{D}}_{\pmb{B}_{S'}}),
\end{equation}
and 
\begin{equation}\label{post_Bs}
\pmb{B}_{S}|\pmb{B}_{S'},\pmb{\Sigma},\pmb{Y}, \pmb{X}_{S;S'} \sim \mathcal{N}_{\mathcal{T}}( \tilde{\pmb{M}}_{\pmb{B}_S}, \pmb{\Sigma}\otimes \tilde{\pmb{D}}_{\pmb{B}_{S}}),
\end{equation}
where
\begin{align*}
\tilde{\pmb{M}}_{\pmb{B}_S}& = \tilde{\pmb{D}}_{\pmb{B}_{S}}\left(\pmb{W}-\pmb{X}_{S}^T\pmb{X}_{S'}\pmb{B}_{S'}\right)
\end{align*}
More details about the computation of the conditional posteriors are provided in Appendix \ref{appendixA}.

The next step is then to propose a Bayesian estimation procedure to estimate the parameters of our models, based on the obtained posterior distributions and a loss function. While only the conditional posterior distributions of the regression parameters derived from the joint posterior distributions are tractable, the Bayesian estimation procedure leads us subsequently to use MCMC algorithms. 
Here, we implement a Collapsed Gibbs Sampler algorithm \citep{van,CGS} to draw samples from the obtained conditional posterior distributions. 

\subsection{MCMC algorithm}\label{mcmc}
In order to draw samples from the target posterior distributions, the implementation of the Collapsed Gibbs sampler algorithm is described as follows :

\subsubsection{Univariate case :}
\begin{itemize}
 \item  Initialize $\pmb{\beta}^{(0)}$  using \eqref{rest} and $\sigma^{2(0)}$,
 \item Update $\pmb{\beta}^{(t)}$ and $\sigma^{2(t)}$ for $ t:1,2, ... $ following the below steps:
  \begin{itemize}
 \item[] Step 1: Draw $\sigma^{2(t)}$ from the distribution of $\sigma^2 | Y, \pmb{X}_{S;S'}\sim \mathcal{IG}(\tilde{\nu},\tilde{\eta})$,
 \item[] Step 2: Draw $\pmb{\beta}_{S'}^{(t)}$ from the distribution of $\pmb{\beta}_{S'}|\sigma ^{2(t)}, Y, \pmb{X}_{S;S'} \sim \mathcal{N}(\tilde{\mu}_{\pmb{\beta}_{S'}}, \sigma^{2(t)}\tilde{C}_{\pmb{\beta}_{S'}})$,
 \item[] Step 3: Calculate the bound and the mean of $\pmb{\beta}_{S}$ from the following equation:
 \begin{equation*}
 \mathcal{T}^{(t)}= \{ \pmb{\beta}_{S} \in \mathbb{R}^{q} | K - H_{S'}\pmb{\beta}_{S'}^{(t)} \leq H_{S} \pmb{\beta}_S \leq G - H_{S'}\pmb{\beta}_{S'}^{(t)}\big \},
\end{equation*}
$$
\tilde{\mu}_{\pmb{\beta}_S}= \tilde{C}_{\pmb{\beta}_S}(W-\pmb{X}_S^T\pmb{X}_{S'}\pmb{\beta}_{S'}^{(t)})
$$
 \item[] Step 4: Draw $\pmb{\beta}_{S}^{(t)}$ from the distribution of $\pmb{\beta}_{S}|\pmb{\beta}_{S'}^{(t)},\sigma ^{2(t)}, Y, \pmb{X}_{S;S'} \sim \mathcal{N}_{\mathcal{T}^{(t)}}(\tilde{\mu}_{\pmb{\beta}_{S}}, \sigma^{2(t)}\tilde{C}_{\pmb{\beta}_{S}})$.
\end{itemize}
 \end{itemize}
 
\subsubsection{Multivariate case :}
\begin{itemize}
  \item Initialize $\pmb{B}^{(0)}$ using \eqref{eq:system:matrix} and $\pmb{\Sigma}^{(0)}$,
  \item Update $\pmb{B}^{(t)}$ and $\pmb{\Sigma}^{(t)}$ for $t: 1, 2, \cdots$ following the below steps:
  \begin{itemize}
    \item[] Step 1: Drew $\pmb{\Sigma}^{(t)}$ from the distribution of $\mathcal{IW}_k (n+r,\pmb{\mathbb{V}})$.
    \item[] Step 2: Drew $\pmb{B}_{S'}^{(t)}$ from the distribution of $\mathcal{N}(\tilde{\pmb{M}}_{\pmb{B}_{S'}} \pmb{\Sigma}^{(t)}\otimes \tilde{\pmb{D}}_{\pmb{B}_{S'}})$.
    \item[] Step 3: Calculate the posterior mean and the restriction inequality bounds  of $vec(\pmb{B}_{S}^{(t)})$ from the following equation as the vector form:
    \begin{equation*}
    vec(\tilde{\pmb{M}}_{\pmb{B}_S}^{(t)})= vec \left(\tilde{\pmb{D}}_{\pmb{B}_{S}}\left(\pmb{W}-\pmb{X}_{S}^T\pmb{X}_{S'}\pmb{B}_{S'}^{(t)}\right)\right),
    \end{equation*}
and
    \begin{equation*}
    \mathcal{S}^{(t)}= \{ vec(\pmb{B}_{S})\in \mathbb{R}^{qk} |{vec(\pmb{K} - \pmb{R}_{S'}\pmb{B}_{S'}^{(t)}) \leq (\pmb{I}_k \otimes \pmb{R}_S) vec(\pmb{B}_{S}^{(t)}) \leq vec(\pmb{G} - \pmb{R}_{S'}\pmb{B}_{S'}^{(t)})}\}.
    \end{equation*}

    \item[] Step 4: Drew $vec(\pmb{B}_{S}^{(t)}) $ from the distribution of $\mathcal{N}_{\mathcal{S}^{(t)}}\left( vec(\tilde{\pmb{M}}_{\pmb{B}_S}^{(t)}), \pmb{\Sigma}^{(t)}\otimes \tilde{\pmb{D}}_{\pmb{B}_{S}}\right)$.
  \end{itemize}
\end{itemize}
See \citet{Geweke2,BRS,Geweke3}'s procedures for more details about the random sampling from the truncated Multivariate Normal Distribution in the last step of both algorithms.
%%%%%%%%%%%%%%%%%%%%%%%%%%%%%%%%%%%%%%%%%%%%%%%%%%
\section{Simulation Studies}\label{simulation}
This study aims to evaluate the performance of the suggested method to estimate the regression model parameters.

\subsection{Univariate linear regression model :}
\begin{exemp}\label{ex1}
 In this example, we consider the following linear regression model 
\begin{equation*}
y_i= \beta_1+ \beta_2 x_{i1}+\beta_3 x_{i2}+\beta_4 x_{i3}+\beta_5 x_{i4}+ \varepsilon _i \qquad i: 1, 2, \cdots, n
\end{equation*}
where $  \varepsilon _i ; \quad  i: 1, 2, \cdots, n$ are i.i.d that follows $ \mathcal{N}(0, \sigma^2) $ and we generated the independent variables from the standard normal distribution. 
The values of  the parameters considered to simulate a dataset of size $n=20$ are
\begin{equation}\label{kk}
\pmb{\beta} = (-0.5, 1, -2, 3, 4)^T, \qquad \sigma ^2 = 1.
\end{equation}
We considered two following restrictions:\\
{\bf Restriction 1.}
 \begin{align}\label{rest-num1}
\beta_2+ \beta_ 3 &\leq -0.5 \nonumber \\ 
\beta_2 +\beta_4 -\beta_5 &\leq 0.2 \nonumber\\
\beta_ 3+\beta_5 &\leq 2.2
\end{align}
{\bf Restriction 2.} \begin{align}\label{rest-num}
\beta_ 2+ \beta_ 3 &\leq -0.5 \nonumber \\ 
\beta_ 3 &\leq -1.5 \nonumber\\
\beta_ 4&\geq 2 
\end{align}

Based on the above restrictions, the matrixes $ H $ and $ G $ would be  
\begin{equation*}
H^{[1]} = 
\begin{pmatrix}
0 & 1 & 1 & 0 & 0  \\
0 & 1 &0 & 1 & -1  \\
0 & 0 & 1 & 0 & 1  \\
\end{pmatrix}, \quad G^{[1]} = 
\begin{pmatrix}
-0.5 \\
0.2\\
2.2\\
\end{pmatrix}, \quad H^{[2]} = 
\begin{pmatrix}
0 & 1 & 1 & 0 & 0  \\
0 & 0 & 1 & 0 & 0  \\
0 & 0 & 0 & -1 & 0  \\
\end{pmatrix}, \quad G^{[2]} = 
\begin{pmatrix}
-0.5 \\
-1.5\\
-2\\
\end{pmatrix}
\end{equation*}

where $ H^{[i]} $ and $ G^{[i]} $ are related to the Restriction $i.$ $i=1,2$. We choose the matrixes  $ H_S^{[i]} $ and $ H_{S'}^{[i]} $ for both Restriction 1. and Restriction 2., respectively, as follows
\begin{equation*}
H_S^{[1]}  = 
\begin{pmatrix}
 1 & 0 & 0  \\
0 & 1 & -1  \\
 1 & 0 & 1  \\
\end{pmatrix}, \quad H_{S'}^{[1]} = \begin{pmatrix}
 0 & 1\\
 0 & 1\\
 0 & 0\\
\end{pmatrix}, \quad H_S^{[2]}  = 
\begin{pmatrix}
 1 & 1 & 0  \\
 0 & 1 & 0  \\
 0 & 0 & -1  \\
\end{pmatrix}, \quad H_{S'}^{[2]} = \pmb{0}_{(3,2)}
\end{equation*}
By these definitions, $ \pmb{\beta}_S$ \text{and} $ \pmb{\beta}_{S'}$ would be
\begin{align*}
\pmb{\beta}_S^{[1]} &= (\beta_3, \beta_4, \beta_5 )^T, \quad \pmb{\beta}_{S'}^{[1]}=(\beta_1, \beta_2)^T, \quad \pmb{\beta}_S^{[2]} = (\beta_2, \beta_3, \beta_4)^T,\quad \pmb{\beta}_{S'}^{[2]}=(\beta_1, \beta_5)^T\\
\pmb{X}_S^{[1]} &= \begin{pmatrix}
X_2 & X_3 & X_4
\end{pmatrix},\quad \pmb{X}_{S'}^{[1]}=\begin{pmatrix}
\pmb{1}_n & X_1
\end{pmatrix},\quad\pmb{X}_S^{[2]} = \begin{pmatrix}
X_1 & X_2 & X_3
\end{pmatrix},\quad \pmb{X}_{S'}^{[2]}=\begin{pmatrix}
\pmb{1}_n & X_4
\end{pmatrix}
\end{align*}
$ \pmb{1}_n $ is an $ n $-length vector of ones and $ X_i $ is an $n$-length vector of the observations for $ i $th independent variable. We then consider the following matrixes 

\begin{equation*}
H_G = 
\begin{pmatrix}
1 & 0 & 0 & 0 & 0 \\
0 & 1 & 1 & 0 & 0  \\
0 & 0 & 1 & 0 & 0  \\
0 & 0 & 0 & -1 & 0  \\
0 & 0 & 0 & 0 & 1
\end{pmatrix}, \quad G_G = 
\begin{pmatrix}
+\infty \\
-0.5 \\
-1.5\\
-2\\
+\infty \\
\end{pmatrix}
\end{equation*}
in order to apply the Geweke's method for the {\bf Restriction 2.}  and to compare it with our approach. 
For both cases, we consider then the following prior distributions 
\begin{align}\label{pmr}
\sigma^2 & \sim \mathcal{IG}(3,1),\nonumber\\
 \pmb{\beta}_{S'} | \sigma^2 & \sim \mathcal{N} (\mu_{S'}, \sigma^2 ( \pmb{X}_{S'}^T  \pmb{X}_{S'})^{-1})\nonumber\\
 \pmb{\beta}_{S} | \pmb{\beta}_{S'}, \sigma^2 & \sim \mathcal{N} (\mu_{S}, \sigma^2 ( \pmb{X}_{S}^T  \pmb{X}_{S})^{-1}).I_{\mathscr{M}}(\pmb{\beta}_{s})
\end{align}
\\
where $ \mu_{s'}$ and $ \mu_{s}$ have been determined using the corresponding elements of the ordinary least square estimation of the parameter $ \pmb{\beta} $, 

\begin{equation*}
\hat{\pmb{\beta}}_{OLS}=  ( \pmb{X}^T  \pmb{X})^{-1} \pmb{X}^T Y,
 \end{equation*}
and 
 \begin{equation*}
 \mathscr{M}=\{ \pmb{\beta}_S \in \mathbb{R}^{3} |H_{S} \pmb{\beta}_S \leq G_{BKS} - H_{S'}\pmb{\beta}_{S'}\}.
\end{equation*}

\begin{table}[ht!]
\begin{center}
\scalebox{0.9}{
\begin{tabular}{lc c c c c c cl}
\toprule
& & {\bf Restriction 1.} &&\multicolumn{2}{c}{\bf Restriction 2.}\\
 \cmidrule{3-3} \cmidrule{5-6} 
& &{BKS-method} &&BKS-method & Geweke's method\\
\bottomrule
 \multirow{2}{*}{$\beta_1$}& $\hat{\beta_1}$ &
-0.5016 &&  -0.4984 & -0.4997 \\
 & SE & 0.2545&& 0.2527 & 0.2556 \\
 & & && &\\
\multirow{2}{*}{$\beta_2$}& $\hat{\beta_2}$ & 1.0013 &&
0.9853 & 0.9720 \\
& SE & 0.2614 && 0.2387 & 0.2396 \\
& & &&  & \\
\multirow{2}{*}{$\beta_3$}& $\hat{\beta_3}$ & -2.0002 &&
-2.0301 & -2.0342 \\
& SE & 0.2019 && 0.2205 & 0.2270 \\
 & & && & \\
\multirow{2}{*}{$\beta_4$}& $\hat{\beta_4}$ & 2.9983 &&
3.0018 & 2.9955 \\
& SE & 0.2216 && 0.2630 & 0.2637 \\
& & && & \\
\multirow{2}{*}{$\beta_5$}& $\hat{\beta_5}$ & 4.0025&&
4.0009 & 4.0002 \\
& SE & 0.1831 && 0.2580 & 0.2608 \\
\midrule
 \multirow{2}{*}{$\sigma^2 $}& $\hat{\sigma^2}$
& 0.7576 && 0.7694 & 0.7611 \\
& SE & 0.2559 && 0.2657& 0.2644\\
 \midrule
\multicolumn{2}{c}{$ MSE$} & 0.0557 && 0.0602 & 0.0627\\
\multicolumn{2}{c}{\text{Time}}&- & & 66.54 & 82.64 \\
\bottomrule
\end{tabular}}
\caption{Bayesian estimations of the regression model parameters, denoted by $\hat{\beta}_j$ and $\hat{\sigma}^2 $, and the standard error, SE, obtained by using the BKS-method and Geweke's method. MSEs are computed using \eqref{mse}, and the time is calculated per minute.} \label{tab-rest1}
\end{center}
\end{table}

Based on the square loss function, we computed the posterior estimations using the samples drawn from the Gibbs sampler algorithm described in Section \ref{mcmc} based on $10^4$ iterations. We tested the convergence of the Markov chains using the Heidelberg and Welch diagnostics for both restrictions. We then chose this iteration number (See Appendix \ref{appendix2} that illustrates the adequacy of the generated Markov chains).
We repeated this experiment 20000 times to calculate the standard error and the mean square error. We used the following function to find the mean square error (MSE) of the estimators
 \begin{equation}\label{mse}
\text{MSE}(\hat{\pmb{\beta}}) = \dfrac{1}{m} \sum_{k=1}^{m} \sum_{j=1}^{p} (\hat{\beta}_{jk}- \beta_j^{real})^2
\end{equation}
where $ m $ is the number of the replications and $ \hat{\beta}_{jk}$ and  $\beta_j^{real} $ are respectively the estimation and the real value of the parameter $ \beta_j $ in $ m $-th replication. 

By comparing the results shown in Table~\ref{tab-rest1} with the true values of the vector $\pmb{\beta}$ and $\sigma^2$, \eqref{kk}, we can see that the posterior estimations of the parameters are very close to the true values. We therefore conclude that the Bayesian approach accurately estimates the model parameters, which is highly due to integrating the accurate prior information into the likelihood.
Note that the analyses carried out in this example are based on the assumption of having enough information to determine {\bf Restriction 1. , Restriction 2.} A scenario is to assume, however, that the probability that the restrictions may not have the forms defined in {\bf Restriction 1. , Restriction 2.} is not zero. An example is the case where we have the following restrictions for instance:
\begin{align}\label{rest-num2}
\beta_2+ \beta_ 3 &\leq -0.5 \nonumber \\ 
\beta_2 +\beta_4 -\beta_5 &\leq 0.2 \nonumber\\
\beta_ 3+\beta_5 &\leq 2.2 + \delta
\end{align}
To simplify the comparison of the restricted Bayesian estimator $ (\hat{\pmb{\beta}}_{RE}) $ (computed using \eqref{pmr}) to the unrestricted Bayesian estimator $  (\hat{\pmb{\beta}}_{UN})$ computed based on the following prior modeling,
\begin{equation*}
\sigma^2 \sim \mathcal{IG}(3,1),
\end{equation*}
\begin{equation*}
 \pmb{\beta} | \sigma^2  \sim \mathcal{N} (\hat{ \pmb{\beta}}_{OLS}, \sigma^2 ( \pmb{X}_{S'}^T  \pmb{X}_{S'})^{-1})
 \end{equation*}
we calculated the relative efficiency defined as follows
\begin{equation}\label{re}
\text{RE} = \dfrac{\text{MSE}(\hat{\pmb{\beta}}_{UN})}{\text{MSE}(\hat{\pmb{\beta}}_{RE})}. 
\end{equation}
If $ \delta \in [-1,1] $, this deviation added to the upper bound of the third restriction in \eqref{rest-num1} has two consequences. When $ \delta $ takes value in $ [-0.2,1] $, the real value of $ \beta_ 3+\beta_5 $ (that is $2$) is in the third restriction range \eqref{rest-num2}. However, when $ \delta \in [-1,0.2) $, the third restriction range does not contain the real value of $ \beta_ 3+\beta_5 $. Figure~\ref{fig1} displays a comparison between the restricted Bayesian estimator $ (\hat{\pmb{\beta}}_{RE}) $ and the unrestricted Bayesian estimator $ (\hat{\pmb{\beta}}_{UN}) $ using the relative efficiency for different values of $\delta$. The figure illustrates that the maximum value of the relative efficiency is located in the interval from $ -0.2 $ to $ 0.2 $, and when $\delta$ exceeds the limit $ 0.2 $, the relative efficiency reduces although that interval still contains the true value of $ \beta_ 3+\beta_5 $. 
These results, therefore, show the impact of the prior information that we have on the restrictions and highlight the role of one's use of this information in determining the parameter estimators.
In other words, when the $ \delta \in [-0.2,1] $, the integration of this a priori accurate information into the model by the Bayesian method, improves the parameter estimations. Then the MSE of the restricted Bayesian estimator is consequently more minor than that of the unrestricted Bayesian estimator. In the case where we suppose $\delta \in [-1,0.2) $, the MSE of the unrestricted Bayesian estimator is less than that of the restricted Bayesian estimator because of the integration of false information in the restriction. 

In addition to this, for the {\bf Restriction 2.}, the results shown in Table~\ref{tab-rest1} display that despite a slight difference between the standard errors and MSE of both methods of estimating, the computing time for the BKS-method is significantly less than that of the Geweke's process.

\begin{figure}[hbt!]
\centering{
\includegraphics[width=0.6\textwidth]{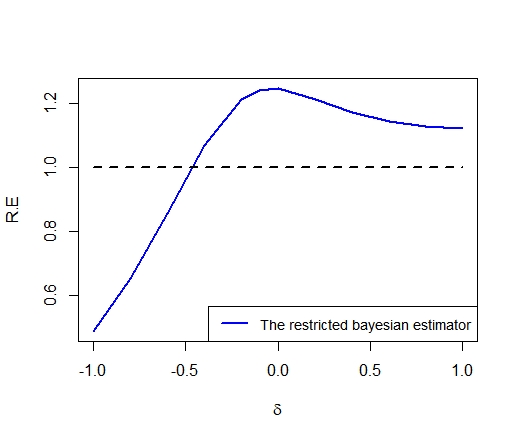}}
\caption{Relative efficiency of the restricted Bayesian estimator for different values of $\delta$.}\label{fig1}
\end{figure}

\end{exemp}

\subsection{Multivariate general linear regression model :}
\begin{exemp}\label{ex2}
%{\color{red} Why do you compare the restricted and un restricted Bayesian methods? I don't see a significant difference between the results presented in table 2. I mean, it seems that, the unrestricted method is as efficient as the restricted Bayesian method even if we add the restriction inequality prior information. So why you complicate your method by adding this restriction prior information then?}

We consider the following multivariate general linear model :
\begin{equation}
  y_{ij}= \beta_{1j}+\beta_{2j}x_{i1}+\beta_{3j}x_{i2}+\beta_{4j}x_{i3}+\beta_{5j}x_{i4}+e_{ij}; \quad i:1, 2, \cdots, 20 \quad \text{and}\quad  j:1, 2
\end{equation}
The true value of $\pmb{B}$ and $\pmb{\Sigma}$ are :
\begin{align*}
\pmb{B} =
\begin{pmatrix}
2 &-1&0.5&1&0.5\\
-1 &1.5&1&1&0.7
\end{pmatrix}^T, \quad
\pmb{\Sigma}= \begin{pmatrix}
1 & 0.5 \\
0.5 & 1
\end{pmatrix}.
\end{align*}
We consider following matrixes for $\pmb{R}$ and $\pmb{G}$:
\begin{align*}
\pmb{R} =
\begin{pmatrix}
0& 1& 0& 0& 0\\
0& 1& 0& 1& 0\\
0& 0& 1& -1& 1
\end{pmatrix}, \quad
\pmb{G}= \begin{pmatrix}
0 & 0 \\
0.5 & 0 \\
0.5 & 1
\end{pmatrix}.
\end{align*}
and the partitions of $\pmb{R}$ are :
\begin{align*}
\pmb{R}_S =
\begin{pmatrix}
0& 0& 0\\
0& 1& 0\\
1& -1& 1
\end{pmatrix}, \quad
\pmb{R}_{S'}= \begin{pmatrix}
0& 1 \\
0& 1 \\
0& 0
\end{pmatrix}
\end{align*}
By these partitions, $\pmb{B}_S$ and $\pmb{B}_{S'}$ would be
\begin{align*}
\pmb{B}_S =
\begin{pmatrix}
\beta_{31}& \beta_{32} \\
\beta_{41}& \beta_{42} \\
\beta_{51}& \beta_{52}
\end{pmatrix}, \quad
\pmb{B}_{S'}= \begin{pmatrix}
\beta_{11}& \beta_{12} \\
\beta_{21}& \beta_{22}
\end{pmatrix}
\end{align*}
The corresponding partitions of the model matrix, $\pmb{X}$, are
\begin{equation*}
\pmb{X}_S = \begin{pmatrix}
X_2 & X_3 & X_4
\end{pmatrix} \qquad  \text{and} \qquad \pmb{X}_{S'}=\begin{pmatrix}
\pmb{1}_n & X_1
\end{pmatrix}
\end{equation*}
where $ \pmb{1}_n $ is a $ n $-length vector of ones and $ X_i $ is the simulated $ i $th independent variable.
We consider then the following prior distributions for the parameters
\begin{equation*}
\pmb{\Sigma} \sim \mathcal{IW}_{k}(2,\pmb{Q}),\qquad \pmb{Q}= \dfrac{1}{20}(\pmb{Y}- \pmb{X}\hat{\pmb{B}}_{OLS})^T (\pmb{Y}- \pmb{X}\hat{\pmb{B}}_{OLS})
\end{equation*}
\begin{equation*}
 \pmb{B}_{S'} | \pmb{\Sigma}  \sim \mathcal{N} \left(\pmb{M}_{\pmb{B}_{S'}}, \pmb{\Sigma} \otimes( \pmb{X}_{S'}^T  \pmb{X}_{S'})^{-1}\right),
 \end{equation*}
 and
 \begin{equation*}
 \pmb{B}_{S} | \pmb{B}_{S'}, \pmb{\Sigma}  \sim \mathcal{N}_{\mathscr{M}} \left(\pmb{M}_{\pmb{B}_S}, \pmb{\Sigma} \otimes ( \pmb{X}_{S}^T  \pmb{X}_{S})^{-1}\right)
\end{equation*}
where $\pmb{M}_{\pmb{B}_{S'}}$ and $\pmb{M}_{\pmb{B}_S}$ are chosen from $\hat{\pmb{B}}_{OLS}$ and
 \begin{equation*}
 \mathscr{M}=\{ \pmb{B}_S \in \mathbb{R}^{3\times 2} | \pmb{R}_S \pmb{B}_{S} \leq \pmb{G} - \pmb{R}_{S'}\pmb{B}_{S'}\}.
\end{equation*}

We use the \textbf{CholWishart} and \textbf{MixMatrix} packages in R programming language, respectively, written by \cite{Cholwishart} and \cite{Mixmatrix}, to draw random samples from the Inverse Wishart and Matrix normal distribution.

Based on the square loss function and the algorithm explained in subsection \ref{mcmc}, we compute the Bayesian parameter estimations using $10^4$ MCMC sample draws. Appendix \ref{appendixB} displays the related trace plots and the sample ACF plots for the regression parameters, which illustrates the convergence of the produced Markov chains. In order to compute the standard error and the mean square error, we repeate this experiment $2000$ times. We use the following function to calculate the MSE of the estimators
 \begin{equation}\label{MSE}
MSE (\hat{\pmb{B}}) = \frac{1}{2000} \sum _{m=1}^{2000} \sum _{i=1}^{p} \sum _{j=1}^{k} \left(\hat{\beta}_{ij(m)}^* - \beta_{ij}^{real}\right)^2,
\end{equation}
where $\hat{\beta}_{ij(m)} $ is the estimation of $\beta_{ij}$ in the $m$th repetition.

\begin{table}[ht!]
\begin{center}
\begin{tabular}{lc c c c c cl}
\toprule
&Parameters & &Estimates& SE\\
\bottomrule
\multirow{5}{*}{$Y_1$}&$\beta_{11}$& & 2.0015 & 0.2527\\
&$\beta_{21}$& & -1.0029 & 0.2499\\
&$\beta_{31}$& & 0.4990 & 0.2249\\
&$\beta_{41}$& & 0.9997 & 0.2381\\
&$\beta_{51}$& & 0.5043 & 0.2215\\
&& & & \\
\multirow{5}{*}{$Y_{2}$}&$\beta_{12}$& & -1.0017 & 0.2510\\
&$\beta_{22}$& & -1.5059 & 0.2529\\
&$\beta_{32}$& & 0.9908 & 0.2149\\
&$\beta_{42}$& & 0.9989 & 0.2230\\
&$\beta_{52}$& & 0.7024 & 0.1952\\
 \midrule
\multicolumn{2}{c}{$ MSE $} && \multicolumn{2}{c}{0.5603}\\
\bottomrule
\multirow{4}{*}{$\pmb{\Sigma}$}& $\sigma_{11} $ & & 0.8859 & 0.2545\\
& $\sigma_{12} $ & & 0.4429 & 0.1973\\
& $\sigma_{21} $ & & 0.4429 & 0.1973\\
& $\sigma_{22} $ & & 0.8830 & 0.2471\\
 \midrule
\multicolumn{2}{c}{$ MSE $} & & \multicolumn{2}{c}{0.3321} \\
\bottomrule
\end{tabular}
\caption{Bayesian estimations of the multivariate regression parameters, $\pmb{B}$ and $\pmb{\Sigma}$, and the standard errors, SE, obtained based on $10^4$ Gibbs iterations (algorithm described in Subsection \ref{mcmc}). The MSEs are computed using \eqref{MSE}.} \label{tab1}
\end{center}
\end{table}
The simulation results shown in Table \ref{tab1} illustrate that the posterior estimations are very close to the true values we considered to simulate our dataset. Thus, our restricted Bayesian approach efficiently estimates the parameters of the multivariate regression model $\pmb{B}$ and $\pmb{\Sigma}$.
\end{exemp}
%%%%%%%%%%%%%%%%%%%

\section{Application to Real Data} \label{real.study}
\subsection{Real dataset 1. Bayesian analysis by applying a univariate linear regression model :}
Following the illustration provided by \cite{Pindyck}, pp. $44$, the real dataset we consider here, contains $32$ observations on rent paid, number of occupants, number of rooms rented, distance from campus in blocks and sex for undergraduates at the University of Michigan. The model considered by \cite{Geweke1, Geweke3} is
\begin{equation}\label{mo}
y_i= \beta _1+ \beta _2 s_i r_i + \beta _3 (1-s_i)r_i+ \beta _4 s_i d_i+ \beta _5 (1-s_i) d_i+ \epsilon_i \qquad i:1, 2, \cdots, 32
\end{equation}
where $ y_i $ denotes rent paid per person, $s_i$ is a dummy variable representing gender (one for male and zero for female), $r_i$ number of rooms per person, $d_i$ distance from campus in blocks, $ \epsilon_i $ is normally distributed error with mean $0$ and variance $ \sigma^2 $. 
The restrictions considered on the parameter $ \beta$ are  
\begin{equation}\label{rest_real}
 \beta _2\geq 0,  \beta_3 \geq 0,  \beta _4 \leq 0, \beta_5 \leq 0  
\end{equation}

In the BKS-method, we can write the restrictions in the form of $ H_{BKS} \pmb{\beta} \leq G_{BKS} $ when the matrix $ H_{BKS} $ and the vector  $ G_{BKS} $ are as follows 

\begin{equation}
H_{BKS}= 
\begin{pmatrix}
0 & -1 & 0 & 0 & 0  \\
0 & 0 & -1 & 0 & 0  \\
0 & 0 & 0 & 1 & 0  \\
0 & 0 & 0 & 0 & 1  \\
\end{pmatrix}, \quad G_{BKS} = 
\begin{pmatrix}
0 \\
0 \\
0\\
0
\end{pmatrix}
\end{equation}
and therefore, we set $H_S$ and $H_{S'}$ as

\begin{equation}
 H_{S} = 
\begin{pmatrix}
 -1 & 0 & 0 & 0  \\
 0 & -1 & 0 & 0  \\
 0 & 0 & 1 & 0  \\
 0 & 0 & 0 & 1  \\
\end{pmatrix}, \quad H_{S'} = 
\begin{pmatrix}
0 \\
0 \\
0\\
0
\end{pmatrix}.
\end{equation}

For the hyperparameters of the prior distributions \eqref{prior1} and \eqref{prior2}, we consider then: 
 \begin{equation}
a = b = 0.001, \quad \mu_{S}= 
\begin{pmatrix}
130.0 \\
123.0 \\
0.0\\
-1.153
\end{pmatrix}, \quad \mu_{S'} = 37.63, \quad  C_{S}= (X_{S}^T X_{S})^{-1}, \quad C_{S'}= (X_{S'}^T X_{S'})^{-1}
\end{equation}
where $ \mu_{S} $ and $ \mu_{S'} $ are chosen based on the MLE of $ \pmb{\beta}$ in the model \eqref{mo} (\cite{Geweke1}), $\hat{\pmb{\beta}}_{MLE}= (37.63, 130.0, 123.0, 0.0, -1.153)^T$. We also implement Geweke's method by considering 
\begin{equation}
H_G= 
\begin{pmatrix}
1 & 0 & 0 & 0 & 0  \\
0 & -1 & 0 & 0 & 0  \\
0 & 0 & -1 & 0 & 0  \\
0 & 0 & 0 & 1 & 0  \\
0 & 0 & 0 & 0 & 1  \\
\end{pmatrix}, \quad G_G = 
\begin{pmatrix}
+\infty \\
0 \\
0 \\
0\\
0
\end{pmatrix}, \quad \mu_{\beta}=\hat{\pmb{\beta}}_{MLE} \quad C_{\beta}= (X^T X)^{-1}.
\end{equation}

\normalsize
\begin{table}[hbt!]
\begin{center}
\begin{tabular}{lc c c cl}
\toprule
Parameters&BKS-method&&  Geweke's method\\
\bottomrule
$\beta_1$& 37.7037 (5.1998) && 37.8017 (21.9386) \\
$\beta_2$& 134.8952 (9.8633) && 134.9228 (24.4277) \\
$\beta_3$& 122.7444 (9.7056) && 122.7571 (25.3568) \\
$\beta_4$& -0.6447 (0.5692) && -0.6573  (0.5787) \\
$\beta_5$& -1.1448 (0.3872) &&  -1.1522 (0.3938) \\
 \midrule
$\sigma^2 $& 1316.165 (349.1228)&& 1323.586 (353.8908)  \\
\bottomrule
\end{tabular}
\caption{Bayesian estimations and corresponding standard deviations (number displayed in brackets) obtained by using the BKS-method and Geweke's method based on $10^4$ MCMC iterations.}\label{tab-real}
\end{center}
\end{table}
The results presented in Table~\ref{tab-real} show that the estimation of the parameters in both Geweke's method and BKS-method are slightly different while the standard deviation of the BKS-method estimates is substantially smaller than that of Geweke's method.

%%%%%%%%%%%%%%%%%%%%%%%%%%%%%%%%%%%%%%%%%%%%%%%%%%%
\subsection{Real dataset 2. Bayesian analysis by applying a multivariate linear regression model :}
As an example of the multivariate general linear model, we consider the chemical reaction data introduced by \cite{Box} and assessed by \cite{Rencher} and \cite{Fujikoshi}.
The explanatory variables are: temperature $(X_1)$, concentration $(X_2)$, time $(X_3)$ and the response variables are: percentage of unchanged starting material $(Y_1)$, percentage converted to the desired product $(Y_2)$, percentage of unwanted by-product $(Y_3)$. 
%%%%%%%%%%%%%%%%%%%%%%%%%%%%%%%%%%%%%%%%%%%%%%%%%%
\begin{comment}
{\color{red}If we denote the matrix of the regression coefficients by $\pmb{B}=\begin{pmatrix}
B_1 \\
B_2 \\
B_3 \\
B_4
\end{pmatrix}$, then multivariate general linear regression model can be defined as follows
\begin{align*}
\pmb{Y}&=\pmb{X}\pmb{B}+\pmb{E}\\
\begin{pmatrix}
Y_1 & Y_2 & Y_3
\end{pmatrix}&=\begin{pmatrix}
\pmb{1}_n & X_1 & X_2 & X_3
\end{pmatrix}\begin{pmatrix}
B_1\\
B_2\\
B_3\\
B_4
\end{pmatrix}+\pmb{E}
\end{align*}
and so for $i$th observation, we have
\begin{equation*}
 \begin{pmatrix}
Y_{i1} & Y_{i2} & Y_{i3}
\end{pmatrix}=\begin{pmatrix}
\beta_{11} & \beta_{12} & \beta_{13}
\end{pmatrix}+x_{i1}\begin{pmatrix}
\beta_{21} & \beta_{22} & \beta_{23}
\end{pmatrix}+x_{i2}\begin{pmatrix}
\beta_{31} & \beta_{32} & \beta_{33}
\end{pmatrix}+x_{i3}\begin{pmatrix}
\beta_{41} & \beta_{42} & \beta_{43}
\end{pmatrix}+\begin{pmatrix}
\epsilon_{i1} & \epsilon_{i2} & \epsilon_{i3}
\end{pmatrix}
\end{equation*}
where $B_l=(\beta_{l1}, \beta_{l2}, \beta_{l3})$ for $l=1, \ldots, 4$, $Y_j$ is a vector of length $n$, $X_k; k=1,2,3$ is an $n-$length vector.}
\end{comment}
%%%%%%%%%%%%%%%%%%%%%%%%%%%%%%%%%%%%%%%%%%%%%%%%%%
We set the following matrixes for defining the restrictions on the coefficient parameters:
\begin{align*}
\pmb{R} =
\begin{pmatrix}
0& 1& 1& -1\\
0& 0& 0& 1\\
\end{pmatrix}, \quad
\pmb{G}= \begin{pmatrix}
-1 & 0.6 & 1.0\\
-2 & 1.5 & 1.5
\end{pmatrix}.
\end{align*}
and then choose the partitions of $\pmb{R}$ as:
\begin{align*}
\pmb{R}_S =
\begin{pmatrix}
1& -1\\
0& 1
\end{pmatrix}, \quad
\pmb{R}_{S'}= \begin{pmatrix}
0& 1 \\
0& 0
\end{pmatrix}.
\end{align*}
By these partitions, $\pmb{B}_S$ and $\pmb{B}_{S'}$ would be
\begin{align*}
\pmb{B}_S =
\begin{pmatrix}
\beta_{31}& \beta_{32} & \beta_{33}\\
\beta_{41}& \beta_{42} & \beta_{43}
\end{pmatrix}, \quad
\pmb{B}_{S'}= \begin{pmatrix}
\beta_{11}& \beta_{12} & \beta_{13}\\
\beta_{21}& \beta_{22} & \beta_{23}
\end{pmatrix}.
\end{align*}
The corresponding partitions of the model matrix $\pmb{X}$ are
\begin{equation*}
\pmb{X}_S = \begin{pmatrix}
X_2 & X_3
\end{pmatrix} \qquad  \text{and} \qquad \pmb{X}_{S'}=\begin{pmatrix}
\pmb{1}_n & X_1
\end{pmatrix}
\end{equation*}
where $ \pmb{1}_n $ is an $ n $-length vector of ones and $ X_i $ is the $ i $th independent variable.
We then specify the hyperparameters of the prior distributions as follows :
\begin{equation*}
\pmb{\Sigma} \sim \mathcal{IW}_{k}(3,\pmb{Q}),\qquad \pmb{Q}= \dfrac{1}{19}(\pmb{Y}- \pmb{X}\hat{\pmb{B}}_{OLS})^T (\pmb{Y}- \pmb{X}\hat{\pmb{B}}_{OLS})
\end{equation*}
\begin{equation*}
 \pmb{B}_{S'} | \pmb{\Sigma}  \sim \mathcal{N}\left(\pmb{M}_{\pmb{B}_{S'}}, \pmb{\Sigma} \otimes( \pmb{X}_{S'}^T  \pmb{X}_{S'})^{-1}\right),
 \end{equation*}
 and
 \begin{equation*}
 \pmb{B}_{S} | \pmb{B}_{S'}, \pmb{\Sigma}  \sim \mathcal{N}_{\mathscr{M}} \left( \pmb{M}_{\pmb{B}_S}, \pmb{\Sigma} \otimes ( \pmb{X}_{S}^T  \pmb{X}_{S})^{-1}\right)
\end{equation*}
where $\pmb{M}_{\pmb{B}_{S'}}$ and $\pmb{M}_{\pmb{B}_S}$ are two matrixes of size $2\times 3$ derived from the MLE, $\hat{\pmb{B}}_{OLS}$, and
 \begin{equation*}
 \mathscr{M}=\{ \pmb{B}_S \in \mathbb{R}^{2\times 3} | \pmb{R}_S \pmb{B}_{S} \leq \pmb{G} - \pmb{R}_{S'}\pmb{B}_{S'}\}.
\end{equation*}

We computed the posterior means of the model parameters and their standard deviations using the samples of size $10^4$ drawn by implementing the Gibbs sampler algorithm described in subsection \ref{mcmc}. Table \ref{tab2} displays the simulation results including the point estimations of the model parameters and their standard deviations.

\begin{table}[ht!]
\begin{center}
\begin{tabular}{lc c c c c cl}
\toprule
&Parameters &&Estimates& Sd\\
\bottomrule
\multirow{5}{*}{$Y_1$}&$\beta_{11}$ && 332.1186 &9.1182\\
&$\beta_{21}$&& -1.5460 & 0.0544 \\
&$\beta_{31}$&& -1.5369 & 0.0974\\
&$\beta_{41}$&& -2.0097 & 0.0107 \\
& && & \\
\multirow{5}{*}{$Y_2$}&$\beta_{12}$&& -25.9678 & 15.8022 \\
&$\beta_{22}$&& 0.4041 & 0.0944 \\
&$\beta_{32}$&& 0.2401 & 0.1441 \\
&$\beta_{42}$&& 1.2626 & 0.0940 \\
& && & \\
\multirow{5}{*}{$Y_3$}& $\beta_{13}$ && -164.1275 & 15.2907 \\
& $\beta_{23}$ && 0.9142 & 0.0913 \\
& $\beta_{33}$ && 0.9404  & 0.1213 \\
& $\beta_{43}$ && 1.2491 & 0.1123 \\
 \midrule
\multirow{6}{*}{$ \pmb{\Sigma}$}& $ \sigma_{11} $ && 4.0024 & 1.2713 \\
& $ \sigma_{12} $&& -1.0698 & 1.5922 \\
& $\sigma_{13} $ && -3.3008 & 1.6979 \\
& $\sigma_{22} $ && 12.4211 & 3.9347 \\
& $\sigma_{23} $ && -8.9508 & 3.3254 \\
& $\sigma_{33} $ && 11.5510 & 3.6744 \\
\bottomrule
\end{tabular}
\caption{Bayesian estimations of the regression parameters, $\pmb{B}$ and $\pmb{\Sigma}$, and the standard deviation obtained based on $10^4$ MCMC iterations provided by the implementation of the algorithm described in Subsection \ref{mcmc}.}\label{tab2}
\end{center}
\end{table}

\section*{Conclusion}
This paper focuses on the Bayesian inference of the univariate and multivariate linear regression models in which the coefficient parameters are subject to linear inequality restrictions. For the univariate case, contrary to Geweke's method, our approach is applicable when the number of the linear constraints on the parameters is less than the number of regression coefficient parameters. In other words, the Bayesian method proposed in this paper leads us to estimate the regression parameters in the case where the restriction matrix is a non-square or non-invertible matrix. We have thus partitioned the restriction matrix of the model coefficients into two matrixes, one of which is square and invertible. 
This partitioning strategy allows us to implement a Gibbs sampling algorithm with the number of parameters to estimate less than the algorithm proposed by Geweke sequentially. Therefore, by decreasing the number of the Gibbs steps, we consequently reduced the running time of the MCMC algorithm using our prior modeling.
 We have evaluated the efficiency of the suggested method with some numerical examples. 
Finally, we applied the BKS-approach to two empirical datasets. The results have shown that the standard deviation of the MCMC samples generated by our suggested method is smaller than that simulated by Geweke's method.

\newpage
\bibliographystyle{ims}  
\bibliography{references}

\begin{thebibliography}{47}
\expandafter\ifx\csname natexlab\endcsname\relax\def\natexlab#1{#1}\fi
\expandafter\ifx\csname url\endcsname\relax
  \def\url#1{\texttt{#1}}\fi
\expandafter\ifx\csname urlprefix\endcsname\relax\def\urlprefix{URL }\fi
\providecommand{\eprint}[2][]{\url{#2}}

\bibitem[{Agrell(2019)}]{Agrell}
\textsc{Agrell, C.} (2019).
\newblock Gaussian processes with linear operator inequality constraints.
\newblock \textit{Journal of Machine Learning Research}, \textbf{20} 1--36.

\bibitem[{Ahmed(2014)}]{Ahmed}
\textsc{Ahmed, S.~E.} (2014).
\newblock \textit{Penalty, shrinkage and pretest strategies-variable selection
  and estimation}.
\newblock Heidelberg: Springer.

\bibitem[{Bahadir et~al.(2017)Bahadir, Yasin and Ahmed}]{Bahadir}
\textsc{Bahadir, Y.}, \textsc{Yasin, A.} and \textsc{Ahmed, S.~E.} (2017).
\newblock Liu-type shrinkage estimations in linear models.
\newblock \textit{arXiv:1709.01131v1 [math.ST]}.

\bibitem[{Bails and Peppers(1982)}]{Bails}
\textsc{Bails, D.~G.} and \textsc{Peppers, L.~C.} (1982).
\newblock \textit{Business Fluctuations}.
\newblock Englewood Cliffs: Prentice-Hall.

\bibitem[{Box and Youle(1995)}]{Box}
\textsc{Box, G.} and \textsc{Youle, P.~V.} (1995).
\newblock The exploration of response surfaces: an example of the link between
  the fitted surface and the basic mechanism of the system.
\newblock \textit{Biometrics}, \textbf{11} 287--323.

\bibitem[{Breslaw(1994)}]{BRS}
\textsc{Breslaw, J.~A.} (1994).
\newblock Random sampling from a truncated multivariate normal distribution.
\newblock \textit{Applied Mathematics Letters}, \textbf{7} 1--6.

\bibitem[{Chamberlain and Leamer(1976)}]{Chamberlain}
\textsc{Chamberlain, G.} and \textsc{Leamer, E.} (1976).
\newblock Matrix weighted averages and posterior bounds.
\newblock \textit{Journal of the Royal Statistical Society, Series B},
  \textbf{38} 73--84.

\bibitem[{Chitsaz and Ahmed(2012a)}]{chitsaz1}
\textsc{Chitsaz, S.} and \textsc{Ahmed, S.~E.} (2012a).
\newblock Shrinkage estimation for the regression parameter matrix in
  multivariate regression model.
\newblock \textit{Journal of Statistical Computation and Simulation},
  \textbf{82} 309--323.

\bibitem[{Chitsaz and Ahmed(2012b)}]{chitsaz2}
\textsc{Chitsaz, S.} and \textsc{Ahmed, S.~E.} (2012b).
\newblock An improved estimation in regression parameter matrix in multivariate
  regression model.
\newblock \textit{Communications in Statistics; Theory and Methods},
  \textbf{41} 2305--2320.

\bibitem[{Davis(1978)}]{Davis}
\textsc{Davis, W.~W.} (1978).
\newblock {B}ayesian analysis of the linear model subject to linear inequality
  constraints.
\newblock \textit{Journal of the Americal Statistical Association}, \textbf{73}
  573--579.

\bibitem[{Dyk and Park(2008)}]{van}
\textsc{Dyk, D. A.~V.} and \textsc{Park, T.} (2008).
\newblock Partially collapsed {G}ibbs samplers.
\newblock \textit{Journal of the American Statistical Association},
  \textbf{103} 790--796.

\bibitem[{Ekvall and Jones(2020)}]{CGS}
\textsc{Ekvall, K.~O.} and \textsc{Jones, G.~L.} (2020).
\newblock Convergence analysis of a collapsed {G}ibbs sampler for {B}ayesian
  vector autoregressions.
\newblock \textit{arXiv preprint arXiv:1907.03170}.

\bibitem[{Escobar and Skarpness(1986)}]{escobar1}
\textsc{Escobar, L.~A.} and \textsc{Skarpness, B.} (1986).
\newblock The bias of the least squares estimator over interval constraints.
\newblock \textit{Economics Letters}, \textbf{20} 331--335.

\bibitem[{Escobar and Skarpness(1987)}]{escobar2}
\textsc{Escobar, L.~A.} and \textsc{Skarpness, B.} (1987).
\newblock Mean square error and efficiency of the least squares estimator over
  interval constraints.
\newblock \textit{Communications in Statistics}, \textbf{16} 397--406.

\bibitem[{Fonseca et~al.(2015)Fonseca, Mexia, Sinha and Zmyslony}]{Fonseca}
\textsc{Fonseca, M.}, \textsc{Mexia, J.~T.}, \textsc{Sinha, B.~K.} and
  \textsc{Zmyslony, R.} (2015).
\newblock Likelihood ratio tests in linear models with linear inequality
  restrictions on regression coefficients.
\newblock \textit{Revstat-Statistical Journal}, \textbf{13} 103--118.

\bibitem[{Fujikoshi et~al.(2010)Fujikoshi, Ulyanov and Shimizu}]{Fujikoshi}
\textsc{Fujikoshi, Y.}, \textsc{Ulyanov, V.~V.} and \textsc{Shimizu, R.}
  (2010).
\newblock \textit{Multivariate Statistics: High-Dimensional and Large-Sample
  Approximations}.
\newblock John Wiley and Sons, Inc., Hoboken, New Jersey.

\bibitem[{Geng and Wan(2000)}]{Geng}
\textsc{Geng, W.} and \textsc{Wan, A. T.~K.} (2000).
\newblock On the sampling performance of an inequality pre-test estimator of
  the regression error variance under {LINEX} loss.
\newblock \textit{Statistical Papers}, \textbf{41} 453--472.

\bibitem[{Geweke(1986)}]{Geweke1}
\textsc{Geweke, J.} (1986).
\newblock Exact inference in the inequality constrained normal linear
  regression model.
\newblock \textit{Journal of Applied Econometrics}, \textbf{1} 127--141.

\bibitem[{Geweke(1991)}]{Geweke2}
\textsc{Geweke, J.} (1991).
\newblock Efficient simulation from the multivariate normal and student
  t-distributions subject to linear constraints.
\newblock \textit{Computer Sciences and Statistics Proceedings of the 23d
  Symposium on the Interface} 571--578.

\bibitem[{Geweke(1996)}]{Geweke3}
\textsc{Geweke, J.} (1996).
\newblock Bayesian inference for linear models subject to linear inequality
  constraints.
\newblock \textit{Modeling and Prediction: Honouring Seymour Geisser, eds. W.
  O. Johnson, J. C. Lee, and A. Zellner, New York, Springer} 248--263.

\bibitem[{Gourieroux et~al.(1982)Gourieroux, Holly and Monfort}]{Gourieroux}
\textsc{Gourieroux, C.}, \textsc{Holly, A.} and \textsc{Monfort, A.} (1982).
\newblock Likelihood ratio test, {W}ald test, and {K}uhn-{T}ucker test in
  linear models with inequality constraints on the regression parameters.
\newblock \textit{Econometrica}, \textbf{50} 63--80.

\bibitem[{Horrace(2005)}]{Horrace}
\textsc{Horrace, W.~C.} (2005).
\newblock Some results on the multivariate truncated normal distribution.
\newblock \textit{Journal of Multivariate Analysis}, \textbf{94} 209--221.

\bibitem[{Izenman(2008)}]{izenman}
\textsc{Izenman, A.~J.} (2008).
\newblock \textit{Modern multivariate statistical techniques: regression,
  classification and manifold learning}.
\newblock New York: Springer.

\bibitem[{Judge and Takayama(1966)}]{Judge}
\textsc{Judge, G.~C.} and \textsc{Takayama, T.} (1966).
\newblock Inequality restrictions in regression analysis.
\newblock \textit{Journal of the American Statistical Association}, \textbf{61}
  166--181.

\bibitem[{Kim and Timm(2007)}]{kim1}
\textsc{Kim, K.} and \textsc{Timm, N.} (2007).
\newblock \textit{Univariate and multivariate general linear models: Theory and
  applications with {SAS}}.
\newblock Second edition, Chapman \& Hall.

\bibitem[{Kim et~al.(2009)Kim, Sohn and Xing}]{kim2}
\textsc{Kim, S.}, \textsc{Sohn, K.~A.} and \textsc{Xing, E.~P.} (2009).
\newblock A multivariate regression approach to association analysis of a
  quantitative trait network.
\newblock \textit{Bioinformatics}, \textbf{25} i204--i212.

\bibitem[{Leamer and Chamberlain(1976)}]{Leamer}
\textsc{Leamer, E.} and \textsc{Chamberlain, G.} (1976).
\newblock A {B}ayesian interpretation of pretesting.
\newblock \textit{Journal of the Royal Statistical Society, Series B},
  \textbf{38} 85--94.

\bibitem[{Lee and Liu(2012)}]{lee}
\textsc{Lee, W.} and \textsc{Liu, Y.} (2012).
\newblock Simultaneous multiple response regression and inverse covariance
  matrix estimation via penalized {G}aussian maximum likelihood.
\newblock \textit{Journal of multivariate analysis}, \textbf{111} 241--255.

\bibitem[{Manolakis and Shaw(2002)}]{Manolakis}
\textsc{Manolakis, D.} and \textsc{Shaw, G.} (2002).
\newblock Detection algorithms for hyperspectral imaging applications.
\newblock \textit{IEEE Signal Processing Magazine}, \textbf{19} 29--43.

\bibitem[{Meng et~al.(2014)Meng, Kuster, Culhane and Gholami}]{meng}
\textsc{Meng, C.}, \textsc{Kuster, B.}, \textsc{Culhane, A.~C.} and
  \textsc{Gholami, A.~M.} (2014).
\newblock A multivariate approach to the integration of multi-omics datasets.
\newblock \textit{BMC Bioinformatics}, \textbf{15}.

\bibitem[{Ohtani(1987)}]{ohtani}
\textsc{Ohtani, K.} (1987).
\newblock The {MSE} of the least squares estimator over an interval constraint.
\newblock \textit{Economics Letters}, \textbf{25} 351--354.

\bibitem[{Pindyck and Rubinfeld(1981)}]{Pindyck}
\textsc{Pindyck, R.~S.} and \textsc{Rubinfeld, D.~L.} (1981).
\newblock \textit{Econometric Models and Economic Forecasts (2nd ed.)}.
\newblock McGraw-Hill: New York.

\bibitem[{Rencher(2002)}]{Rencher}
\textsc{Rencher, A.~C.} (2002).
\newblock \textit{Methods of multivariate analysis}.
\newblock Second edition, Wiley, Hoboken, NJ.

\bibitem[{Rodriguez-Yam et~al.(2004)Rodriguez-Yam, Davis and
  Scharf}]{Rodriguez2}
\textsc{Rodriguez-Yam, G.}, \textsc{Davis, R.~A.} and \textsc{Scharf, L.~L.}
  (2004).
\newblock Efficient gibbs sampling of truncated multivariate normal with
  application to constrained linear regression.
\newblock In \textit{Technical report, Colorado State University}. Unpublished
  manuscript.

\bibitem[{Rodriguez-Yam et~al.(2002)Rodriguez-Yam, Davis and
  Scharf}]{Rodriguez1}
\textsc{Rodriguez-Yam, G.~A.}, \textsc{Davis, R.~A.} and \textsc{Scharf, L.~L.}
  (2002).
\newblock A {B}ayesian model and {G}ibbs sampler for hyperspectral imaging.
\newblock \textit{In Proceedings of the 2002 IEEE Sensor Array and Multichannel
  Signal Processing Workshop, Washington, D. C.} 105--109.

\bibitem[{Rowe(2003)}]{Rowe}
\textsc{Rowe, D.~B.} (2003).
\newblock \textit{Multivariate {B}ayesian statistics: models for source
  separation and signal unmixing}.
\newblock Chapman \& Hall/CRC.

\bibitem[{Srivastava and T.~K.~Solanky(2003)}]{srivastava}
\textsc{Srivastava, M.~S.} and \textsc{T.~K.~Solanky, T.~K.} (2003).
\newblock Predicting multivariate response in linear regression model.
\newblock \textit{Marcel Dekker. Inc.}, \textbf{32} 389--409.

\bibitem[{Thompson et~al.(2019{\natexlab{a}})Thompson, Ripley, Venables and
  Thompson}]{Mixmatrix}
\textsc{Thompson, G.}, \textsc{Ripley, B.~D.}, \textsc{Venables, W.~N.} and
  \textsc{Thompson, M.~G.} (2019{\natexlab{a}}).
\newblock Package mixmatrix.
\newblock \textit{arXiv: 1907.09565}.

\bibitem[{Thompson et~al.(2019{\natexlab{b}})Thompson, Team and
  Thompson}]{Cholwishart}
\textsc{Thompson, G.}, \textsc{Team, R.~C.} and \textsc{Thompson, M.~G.}
  (2019{\natexlab{b}}).
\newblock Package cholwishart.
\newblock
  \textit{https://cran.r-project.org/web/packages/CholWishart/CholWishart.pdf}.

\bibitem[{Timm(2002)}]{timm}
\textsc{Timm, N.~H.} (2002).
\newblock \textit{Applied multivariate analysis}.
\newblock New York: Springer.

\bibitem[{Veiga and Marrel(2012)}]{Veiga}
\textsc{Veiga, S.~D.} and \textsc{Marrel, A.} (2012).
\newblock Gaussian process modeling with inequality constraints.
\newblock \textit{Annales de la facult{é} des sciences de Toulouse
  Math{é}matiques}, \textbf{21} 529--555.

\bibitem[{Wang(2013)}]{wang}
\textsc{Wang, J.} (2013).
\newblock Joint estimation of sparse multivariate regression and conditional
  graphical models.
\newblock \textit{arXiv preprint arXiv: 1306.4410}.

\bibitem[{Wolak(1987)}]{Wolak1}
\textsc{Wolak, F.} (1987).
\newblock An exact test for multiple inequality and equality constraints in the
  linear model,.
\newblock \textit{Journal of the American Statistical Association}, \textbf{82}
  782--793.

\bibitem[{Wolak(1989)}]{Wolak2}
\textsc{Wolak, F.} (1989).
\newblock Testing inequality constraints in linear econometric models.
\newblock \textit{Journal of Econometrics}, \textbf{41} 205--235.

\bibitem[{Zapala et~al.(2005)Zapala, Hovatta, Ellison, Wodicka, Rio, Tennant,
  Tynan, Broide, Helton, Stoveken, Winrow, Lockhart, Reilly, Young, Bloom,
  Lockhart and Barlow}]{zapala}
\textsc{Zapala, M.~A.}, \textsc{Hovatta, I.}, \textsc{Ellison, J.~A.},
  \textsc{Wodicka, L.}, \textsc{Rio, J. A.~D.}, \textsc{Tennant, R.},
  \textsc{Tynan, W.}, \textsc{Broide, R.~S.}, \textsc{Helton, R.},
  \textsc{Stoveken, B.~S.}, \textsc{Winrow, C.}, \textsc{Lockhart, D.~J.},
  \textsc{Reilly, J.~F.}, \textsc{Young, W.~G.}, \textsc{Bloom, F.~E.},
  \textsc{Lockhart, D.~J.} and \textsc{Barlow, C.} (2005).
\newblock Adult mouse brain gene expression patterns beer an embryologic
  imprint.
\newblock \textit{Proc. Natl Acad. Sci.}, \textbf{102} 10357--10362.

\bibitem[{Zhu et~al.(2005)Zhu, Santerre and Chang}]{Zhu1}
\textsc{Zhu, J.}, \textsc{Santerre, R.} and \textsc{Chang, X.~W.} (2005).
\newblock A {B}ayesian method for linear, inequality constrained adjustment and
  its application to {GPS} positioning.
\newblock \textit{Journal of Geodesy}, \textbf{78} 528--534.

\bibitem[{Zhu and Zhou(2014)}]{ZHU2}
\textsc{Zhu, R.} and \textsc{Zhou, S. Z.~F.} (2014).
\newblock Testing inequality constraints in a linear regression model with
  spherically symmetric disturbances.
\newblock \textit{Journal of Systems Science and Complexity}, \textbf{27}
  1204--1212.

\end{thebibliography}

\newpage
\appendix

\section{Computing the conditional posterior distributions of the restricted regression model \eqref{m12}}
\label{Appendix1}
The likelihood function of the model \eqref{m12} is given by
\begin{align}\label{like1}
\ell(\pmb{\beta}_{S;S'}, \sigma^2 |Y, \pmb{X}_{S;S'}) 
&=\frac{1}{(2\pi)^{\frac{n}{2}}[\sigma^2]^{\frac{n}{2}}}
\exp\left(-\frac{1}{2\sigma^2}(Y-\pmb{X}_{S}\pmb{\beta}_{S}-\pmb{X}_{S'}\pmb{\beta}_{S'} )^T
(Y- \pmb{X}_{S}\pmb{\beta}_{S}-\pmb{X}_{S'}\pmb{\beta}_{S'}) \right)
\end{align}
 By the definition, the joint posterior distribution $\pi(\pmb{\beta}_{S;S'}, \sigma^2|Y, \pmb{X}_{S;S'})$ is then
 \begin{align*}
 \pi(\pmb{\beta}_{S;S'}, \sigma^2 & |Y, \pmb{X}_{S;S'}) = \ell(\pmb{\beta}_{S;S'}, \sigma^2|Y, \pmb{X}_{S;S'}) \pi(\pmb{\beta}_{S}|\pmb{\beta}_{S'}, \sigma^2) \pi(\pmb{\beta}_{S'}|\sigma^2)\pi(\sigma^2)\nonumber\\
&\propto \frac{1}{[\sigma^2]^{\frac{n}{2}}}\exp\left(-\frac{1}{2\sigma^2} (Y-\pmb{X}_{S}\pmb{\beta}_{S}-\pmb{X}_{S'}\pmb{\beta}_{S'} )^T(Y- \pmb{X}_{S}\pmb{\beta}_{S}-\pmb{X}_{S'}\pmb{\beta}_{S'}) \right) \nonumber\\
&\times \frac{1}{[\sigma^2]^{\frac{p-q}{2}}}\exp\left(-\frac{1}{2\sigma^2}(\pmb{\beta}_{S'}- \mu_{\pmb{\beta}_{S'}})^T C_{\pmb{\beta}_{S'}}^{-1} (\pmb{\beta}_{S'}- \mu_{\pmb{\beta}_{S'}}) \right) \nonumber\\
&\times \frac{1}{[\sigma^2]^{\frac{q}{2}}}\exp\left(-\frac{1}{2\sigma^2}(\pmb{\beta}_{S}- \mu_{\pmb{\beta}_{S}})^T C_{\pmb{\beta}_{S}}^{-1}(\pmb{\beta}_{S}- \mu_{\pmb{\beta}_{S}}) \right). I_{\mathcal{T}}  (\pmb{\beta}_S) \nonumber\\
&\times  \frac{1}{[\sigma^2]^{\frac{a}{2}+1}} \exp \left(-\frac{b}{2\sigma^2} \right) \nonumber\\
\end{align*}
After some calculation, we obtain 
\begin{align}\label{post1}
\pi(&\pmb{\beta}_{S;S'}, \sigma^2  | Y, \pmb{X}_{S;S'}) \propto [\sigma^2]^{-\frac{n+a +p}{2}-1} \exp \left(  -\dfrac{1}{2 \sigma ^2} (b+ Y^T Y + \mu_{\pmb{\beta}_{S'}}^{T}C_{\pmb{\beta}_{S'}}^{-1} \mu_{\pmb{\beta}_{S'}} +\mu_{\pmb{\beta}_{S}}^{T}C_{\pmb{\beta}_{S}}^{-1} \mu_{\pmb{\beta}_{S}})  \right) \nonumber\\
& \times  \exp \left( -\dfrac{1}{2 \sigma ^2} (\pmb{\beta}_{S}^{T}  \tilde{C}_{\pmb{\beta}_S}^{-1}\pmb{\beta}_{S}- \pmb{\beta}_{S}^{T}(W- \pmb{X}_{S}^{T} \pmb{X}_{S'} \pmb{\beta}_{S'})- (W- \pmb{X}_{S}^{T} \pmb{X}_{S'} \pmb{\beta}_{S'})^{T} \pmb{\beta}_{S} ) \right).  I_{\mathcal{T}}  (\pmb{\beta}_S)\nonumber\\
& \times \exp \left( -\dfrac{1}{2 \sigma ^2} (\pmb{\beta}_{S'}^{T} (\pmb{X}_{S'}^{T} \pmb{X}_{S'}+ C_{\pmb{\beta}_{S'}}^{-1})\pmb{\beta}_{S'} - \pmb{\beta}_{S'}^{T} (C_{\pmb{\beta}_{S'}}^{-1}\mu_{\pmb{\beta}_{S'}} + \pmb{X}_{S'}^{T} Y) - (C_{\pmb{\beta}_{S'}}^{-1}\mu_{\pmb{\beta}_{S'}} + \pmb{X}_{S'}^{T} Y)^{T} \pmb{\beta}_{S'}) \right)
\end{align}
\normalsize
\\
in which
\begin{align}
 \tilde{C}_{\pmb{\beta}_S}^{-1}&= (\pmb{X}_{S}^{T}\pmb{X}_{S}+ C_{\pmb{\beta}_{S}}^{-1} ) \nonumber\\
W&= \pmb{X}_{S}^{T} Y +  C_{\pmb{\beta}_{S}}^{-1}  \mu_{\pmb{\beta}_{S}} \nonumber
\end{align}
If we suppose that
$$
\tilde{\mu}_{\pmb{\beta}_S}= \tilde{C}_{\pmb{\beta}_S}(W-\pmb{X}_S^T\pmb{X}_{S'}\pmb{\beta}_{S'})
$$
then the second line of the joint posterior distribution \eqref{post1} can be rewritten as follows
\begin{align}\label{post2}
&\exp \left(-\frac{1}{2\sigma^2}(\pmb{\beta}_S-\tilde{\mu}_{\pmb{\beta}_S})^T \tilde{C}_{\pmb{\beta}_S}^{-1}(\pmb{\beta}_S-\tilde{\mu}_{\pmb{\beta}_S}) + \frac{1}{2\sigma^2}(W - \pmb{X}_S^T\pmb{X}_{S'}\pmb{\beta}_{S'})^T \tilde{C}_{\pmb{\beta}_S}(W - \pmb{X}_S^T\pmb{X}_{S'}\pmb{\beta}_{S'}) \right)I_{\mathcal{T}}  (\pmb{\beta}_S)
\end{align}
and by replacing the expression \eqref{post2} in \eqref{post1}, we get
\begin{align}\label{post11}
\pi(&\pmb{\beta}_{S;S'}, \sigma^2  | Y, \pmb{X}_{S;S'}) \propto [\sigma^2]^{-\frac{n+a +p}{2}-1} \exp \left(  -\dfrac{1}{2 \sigma ^2} (b+ Y^T Y + \mu_{\pmb{\beta}_{S'}}^{T}C_{\pmb{\beta}_{S'}}^{-1} \mu_{\pmb{\beta}_{S'}} +\mu_{\pmb{\beta}_{S}}^{T}C_{\pmb{\beta}_{S}}^{-1} \mu_{\pmb{\beta}_{S}})  \right) \nonumber\\
& \times  \exp \left(-\frac{1}{2\sigma^2}(\pmb{\beta}_S-\tilde{\mu}_{\pmb{\beta}_S})^T \tilde{C}_{\pmb{\beta}_S}^{-1}(\pmb{\beta}_S-\tilde{\mu}_{\pmb{\beta}_S}) \right)I_{\mathcal{T}}  (\pmb{\beta}_S)\nonumber\\
& \times \exp\left( \frac{1}{2\sigma^2}(W - \pmb{X}_S^T\pmb{X}_{S'}\pmb{\beta}_{S'})^T \tilde{C}_{\pmb{\beta}_S}(W - \pmb{X}_S^T\pmb{X}_{S'}\pmb{\beta}_{S'}) \right)\nonumber\\
& \times \exp \left( -\dfrac{1}{2 \sigma ^2} (\pmb{\beta}_{S'}^{T} (\pmb{X}_{S'}^{T} \pmb{X}_{S'}+ C_{\pmb{\beta}_{S'}}^{-1})\pmb{\beta}_{S'} - \pmb{\beta}_{S'}^{T} (C_{\pmb{\beta}_{S'}}^{-1}\mu_{\pmb{\beta}_{S'}} + \pmb{X}_{S'}^{T} Y) - (C_{\pmb{\beta}_{S'}}^{-1}\mu_{\pmb{\beta}_{S'}} + \pmb{X}_{S'}^{T} Y)^{T} \pmb{\beta}_{S'}) \right)
\end{align}
\normalsize
and by supposing
\begin{align*}
\tilde{C}_{\pmb{\beta}_{S'}}^{-1}&=\pmb{X}_{S'}^{T} \pmb{X}_{S'}+ C_{\pmb{\beta}_{S'}}^{-1}-(\pmb{X}_S^T\pmb{X}_{S'})^T \tilde{C}_{\pmb{\beta}_S}(\pmb{X}_S^T\pmb{X}_{S'})\\
\tilde{\mu}_{\pmb{\beta}_{S'}}&=\tilde{C}_{\pmb{\beta}_{S'}}(C_{\pmb{\beta}_{S'}}^{-1}\mu_{\pmb{\beta}_{S'}} + \pmb{X}_{S'}^{T} Y-(\pmb{X}_{S'}^T\pmb{X}_S) \tilde{C}_{\pmb{\beta}_S}W)
\end{align*}
the joint posterior distribution \eqref{post11} can then be rewritten as follows
\begin{align}\label{post111}
\pi(&\pmb{\beta}_{S;S'}, \sigma^2  | Y, \pmb{X}_{S;S'}) \propto [\sigma^2]^{-\frac{n+a +p}{2}-1} \exp \left(  -\dfrac{1}{2 \sigma ^2} (b+ Y^T Y + \mu_{\pmb{\beta}_{S'}}^{T}C_{\pmb{\beta}_{S'}}^{-1} \mu_{\pmb{\beta}_{S'}} +\mu_{\pmb{\beta}_{S}}^{T}C_{\pmb{\beta}_{S}}^{-1} \mu_{\pmb{\beta}_{S}})  \right) \nonumber\\
& \times  \exp \left(-\frac{1}{2\sigma^2}(\pmb{\beta}_S-\tilde{\mu}_{\pmb{\beta}_S})^T \tilde{C}_{\pmb{\beta}_S}^{-1}(\pmb{\beta}_S-\tilde{\mu}_{\pmb{\beta}_S}) \right)I_{\mathcal{T}}  (\pmb{\beta}_S)\nonumber\\
&\times \exp \left( -\dfrac{1}{2 \sigma ^2} (\pmb{\beta}_{S'}-\tilde{\mu}_{\pmb{\beta}_{S'}})^T\tilde{C}_{\pmb{\beta}_{S'}}^{-1}(\pmb{\beta}_{S'}-\tilde{\mu}_{\pmb{\beta}_{S'}}) + \dfrac{1}{2 \sigma ^2}(\tilde{\mu}_{\pmb{\beta}_{S'}}^T\tilde{C}_{\pmb{\beta}_{S'}}^{-1}\tilde{\mu}_{\pmb{\beta}_{S'}})+\dfrac{1}{2 \sigma ^2}W^T \tilde{C}_{\pmb{\beta}_S}W\right)
\end{align}
\normalsize
%%%%%%%%%%%%%%%%%%%%%%%%%%%%%%%%%%%%%%%%%%
From the joint posterior distribution \eqref{post111}, we can easily deduce the marginal posterior distributions \eqref{cond1}, \eqref{cond2} and \eqref{cond3}.

\newpage
\section{Appendix: Trace and sample ACF plots of the simulated Markov Chains in section \ref{simulation}}\label{appendix2}
\begin{figure}[hbt!]
\centering{
\includegraphics[width=1\textwidth]{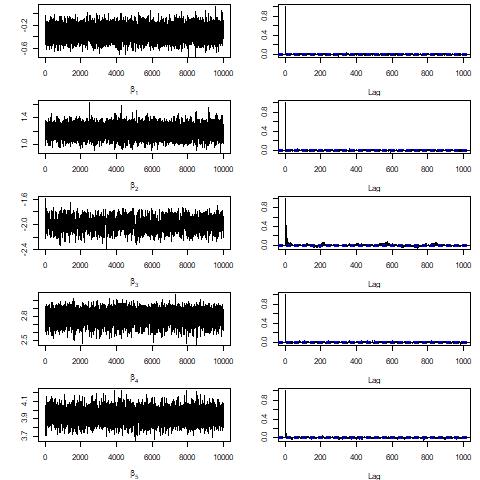}}
\caption{Trace plots and the sample ACF plots of $10^4$ MCMC iterations simulated from the posterior distribution of $\pmb{\beta}$ in the \textbf{Restriction 1.} using the algorithm described in subsection \ref{mcmc}.}\end{figure}

\begin{figure}[hbt!]
\centering{\includegraphics[width=1\textwidth]{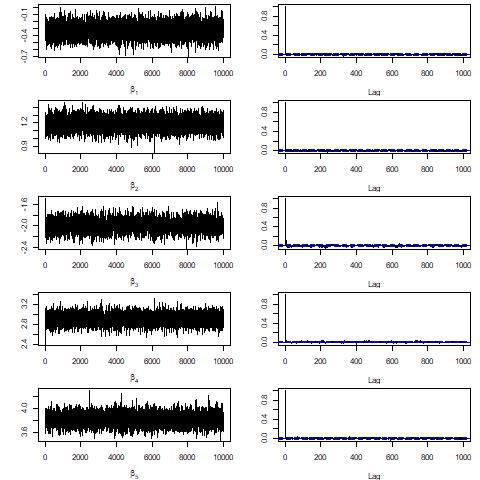}}
\caption{Trace plots and the sample ACF plots of $10^4$ MCMC iterations simulated from posterior distribution of $\pmb{\beta}$ in the \textbf{Restriction 2.} using the algorithm described in subsection \ref{mcmc}.}\end{figure}

%%%%%%%%%%%%%%%%%%%%%%%%%%%%%%%%%%%%%%%%%%
%%%%%%%%%%%%%%%%%%%%%%%%%%%%%%%%%%%%%%%%%%
\newpage
\section{Appendix: Computing the conditional posterior distributions of the regression parameters for the model \eqref{m22}}\label{appendixA}

The likelihood function of the model \eqref{m22} is given by
\begin{align}\label{like21}
 \ell(\pmb{B}_{S;S'},\pmb{\Sigma} |\pmb{Y}, \pmb{X}_{S;S'})& \propto |\pmb{\Sigma}|^{-\frac{n}{2}} 
 \exp \bigg( -\frac{1}{2} tr \big\lbrace \pmb{\Sigma}^{-1}(\pmb{Y}-\pmb{X}_{S}\pmb{B}_{S}-\pmb{X}_{S'}\pmb{B}_{S'})^T
(\pmb{Y}- \pmb{X}_{S}\pmb{B}_{S}-\pmb{X}_{S'}\pmb{B}_{S'})\big\rbrace \bigg)
\end{align}
By the definition, the posterior distribution $\pi(\pmb{\beta}_{S;S'}, \pmb{\Sigma} |Y, \pmb{X}_{S;S'})$ is then
 \begin{align*}
 \pi(\pmb{B}_{S;S'}, \pmb{\Sigma} & |\pmb{Y}, \pmb{X}_{S;S'}) = \ell(\pmb{B}_{S;S'}, \pmb{\Sigma}|\pmb{Y}, \pmb{X}_{S;S'}) \pi(\pmb{B}_{S}|\pmb{B}_{S'}, \pmb{\Sigma}) \pi(\pmb{B}_{S'}|\pmb{\Sigma})\pi(\pmb{\Sigma})\nonumber\\
&\propto |\pmb{\Sigma}|^{-\frac{n}{2}} \exp \bigg( -\frac{1}{2} tr \big\lbrace \pmb{\Sigma}^{-1}(\pmb{Y}-\pmb{X}_{S}\pmb{B}_{S}-\pmb{X}_{S'}\pmb{B}_{S'})^T
(\pmb{Y}- \pmb{X}_{S}\pmb{B}_{S}-\pmb{X}_{S'}\pmb{B}_{S'})\big\rbrace \bigg)\\
& \times |\pmb{\Sigma}|^{-\frac{p-q}{2}}\exp \bigg( -\frac{1}{2} tr \big\lbrace \pmb{\Sigma}^{-1}(\pmb{B}_{S'}-\pmb{M}_{\pmb{B}_{S'}})^T
\pmb{D}_{\pmb{B}_{S'}}^{-1}(\pmb{B}_{S'}-\pmb{M}_{\pmb{B}_{S'}})\big\rbrace \bigg) \\
&\times |\pmb{\Sigma}|^{-\frac{q}{2}}\exp \bigg( -\frac{1}{2} tr \big\lbrace \pmb{\Sigma}^{-1}(\pmb{B}_{S}-\pmb{M}_{\pmb{B}_S})^T
\pmb{D}_{\pmb{B}_S}^{-1}(\pmb{B}_{S}-\pmb{M}_{\pmb{B}_S})\big\rbrace \bigg). I_{\mathcal{T}}(\pmb{B}_S) \nonumber\\
&\times |\pmb{\Sigma}|^{-\frac{r}{2}}\exp \bigg( -\frac{1}{2} tr \big\lbrace \pmb{\Sigma}^{-1} \pmb{Q} \big\rbrace \bigg) \nonumber
\end{align*}
and then, we obtain
\begin{align} \label{post22}
\pi(&\pmb{B}_{S;S'}, \pmb{\Sigma} | \pmb{Y}, \pmb{X}_{S;S'}) \propto |\pmb{\Sigma}|^{-\frac{n+r+p}{2}} \exp \bigg(  -\dfrac{1}{2} tr \big\lbrace \pmb{\Sigma}^{-1} \big(\pmb{Q}+ \pmb{Y}^T \pmb{Y} + \pmb{M}_{\pmb{B}_{S'}}^{T}\pmb{D}_{\pmb{B}_{S'}}^{-1} \pmb{M}_{\pmb{B}_{S'}} +\pmb{M}_{\pmb{B}_S}^{T}\pmb{D}_{\pmb{B}_S}^{-1} \pmb{M}_{\pmb{B}_S})\big\rbrace  \bigg) \nonumber \\
& \times  \exp \bigg( -\dfrac{1}{2}tr \big\lbrace \pmb{\Sigma}^{-1} (\pmb{B}_{S}^{T} \tilde{\pmb{D}}_{\pmb{B}_{S}}^{-1}\pmb{B}_{S}- \pmb{B}_{S}^{T}(\pmb{W}- \pmb{X}_{S}^{T} \pmb{X}_{S'} \pmb{B}_{S'})- (\pmb{W}- \pmb{X}_{S}^{T} \pmb{X}_{S'} \pmb{B}_{S'})^{T} \pmb{B}_{S})\big\rbrace \bigg).  I_{\mathcal{T}}  (\pmb{B}_S) \nonumber\\
& \times \exp \bigg( -\dfrac{1}{2}tr \big\lbrace \pmb{\Sigma}^{-1}(\pmb{B}_{S'}^{T} (\pmb{X}_{S'}^{T} \pmb{X}_{S'}+\pmb{D}_{\pmb{B}_{S'}}^{-1}) \pmb{B}_{S'}-\pmb{X}_{S'}^{T}(\pmb{D}_{\pmb{B}_{S'}}^{-1}\pmb{M}_{\pmb{B}_{S'}} + \pmb{B}_{S'}^{T} \pmb{Y})- (\pmb{D}_{\pmb{B}_{S'}}^{-1}\pmb{M}_{\pmb{B}_{S'}} + \pmb{X}_{S'}^{T} \pmb{Y})^{T} \pmb{B}_{S'}) \big\rbrace \bigg)
\end{align}
\normalsize
in which
\begin{align}
 \tilde{\pmb{D}}_{\pmb{B}_{S}}^{-1}&= (\pmb{X}_{S}^{T}\pmb{X}_{S}+ \pmb{D}_{\pmb{B}_S}^{-1} ) \nonumber\\
\pmb{W}&= \pmb{X}_{S}^{T} \pmb{Y} + \pmb{D}_{\pmb{B}_S}^{-1} \pmb{M}_{\pmb{B}_S}. \nonumber
\end{align}
If we suppose that
\begin{equation}
 \tilde{\pmb{M}}_{\pmb{B}_S}=\tilde{\pmb{D}}_{\pmb{B}_{S}}^{-1}(\pmb{W}- \pmb{X}_{S}^{T} \pmb{X}_{S'} \pmb{B}_{S'})
\end{equation}
then the second line of \eqref{post22} can be written as
\begin{align}\label{post23}
  \exp \bigg( -\dfrac{1}{2}tr \big\lbrace \pmb{\Sigma}^{-1} (\pmb{B}_{S}^{T}-  \tilde{\pmb{M}}_{\pmb{B}_S})^T & \tilde{\pmb{D}}_{\pmb{B}_{S}}^{-1}(\pmb{B}_{S}^{T}-  \tilde{\pmb{M}}_{\pmb{B}_S}) 
 - (\pmb{W}- \pmb{X}_{S}^{T} \pmb{X}_{S'} \pmb{B}_{S'})^{T}\tilde{\pmb{D}}_{\pmb{B}_{S}}(\pmb{W}- \pmb{X}_{S}^{T} \pmb{X}_{S'} \pmb{B}_{S'}) \big\rbrace \bigg).  I_{\mathcal{T}}  (\pmb{B}_S)
\end{align}
and finally, by replacing \eqref{post23} in \eqref{post22} and supposing that
 \begin{align*}
 \tilde{\pmb{D}}_{\pmb{B}_{S'}}^{-1}&= \pmb{X}_{S'}^{T}\pmb{X}_{S'}+ \pmb{D}_{\pmb{B}_{S'}}^{-1}- (\pmb{X}_{S}^{T} \pmb{X}_{S'})^T \tilde{\pmb{D}}_{\pmb{B}_{S}}(\pmb{X}_{S}^{T} \pmb{X}_{S'}) \nonumber\\
  \tilde{\pmb{M}}_{\pmb{B}_{S'}} & = \tilde{\pmb{D}}_{\pmb{B}_{S'}} \big(\pmb{D}_{\pmb{B}_{S'}}^{-1}\pmb{M}_{\pmb{B}_S} + \pmb{X}_{S'}^{T} \pmb{Y} -  \pmb{X}_{S'}^{T}\pmb{X}_{S} \tilde{\pmb{D}}_{\pmb{B}_{S}}\pmb{W}\big)\\
  \pmb{\mathbb{V}} & =\pmb{Q}+ \pmb{Y}^T \pmb{Y} + \pmb{M}_{\pmb{B}_{S'}}^{T}\pmb{D}_{\pmb{B}_{S'}}^{-1} \pmb{M}_{\pmb{B}_{S'}} +\pmb{M}_{\pmb{B}_S}^{T}\pmb{D}_{\pmb{B}_S}^{-1} \pmb{M}_{\pmb{B}_S}
 -\pmb{W}^T \tilde{\pmb{D}}_{\pmb{B}_{S}} \pmb{W} - \tilde{\pmb{M}}_{\pmb{B}_{S'}}^T  \tilde{\pmb{D}}_{\pmb{B}_{S'}}^{-1} \tilde{\pmb{M}}_{\pmb{B}_{S'}}
\end{align*}
the joint posterior distribution will be
 \begin{align*}
 \pi(\pmb{B}_{S;S'}, \pmb{\Sigma} |\pmb{Y}, \pmb{X}_{S;S'}) &\propto |\pmb{\Sigma}|^{-\frac{n+r}{2}} \exp \bigg( -\frac{1}{2} tr \big\lbrace \pmb{\Sigma}^{-1}\pmb{\mathbb{V}} \big\rbrace \bigg)\\
& \times |\pmb{\Sigma}|^{-\frac{q}{2}} \exp \bigg( -\dfrac{1}{2}tr \big\lbrace \pmb{\Sigma}^{-1} (\pmb{B}_{S}^{T}-  \tilde{\pmb{M}}_{\pmb{B}_S})^T \tilde{\pmb{D}}_{\pmb{B}_{S}}^{-1}(\pmb{B}_{S}^{T}-  \tilde{\pmb{M}}_{\pmb{B}_S})\big\rbrace \bigg).  I_{\mathcal{T}}  (\pmb{B}_S) \\
& \times |\pmb{\Sigma}|^{-\frac{p-q}{2}} \exp \bigg( -\dfrac{1}{2}tr \big\lbrace \pmb{\Sigma}^{-1} (\pmb{B}_{S'}^{T}-  \tilde{\pmb{M}}_{\pmb{B}_{S'}})^T \tilde{\pmb{D}}_{\pmb{B}_{S'}}^{-1}(\pmb{B}_{S'}^{T}-  \tilde{\pmb{M}}_{\pmb{B}_{S'}})\big\rbrace \bigg).
\end{align*}
\newpage
\section{Appendix: Trace and sample ACF plots of the simulated Markov Chains in Section \ref{simulation}}\label{appendixB}

\begin{figure}[hbt!]
\centering{
\includegraphics[width=1\textwidth]{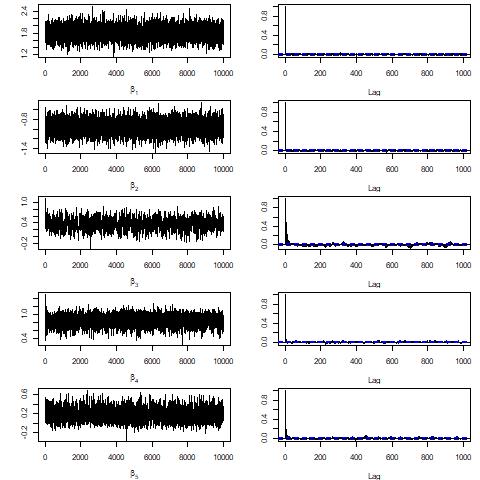}}
\caption{Trace plots and the sample ACF plots of $10^4$ MCMC iterations from posterior distributions using the algorithm described in subsection \ref{mcmc} for coefficient parameters of response variable $Y_1$.}\label{fig1}
\end{figure}

\begin{figure}[hbt!]
\centering{
\includegraphics[width=1\textwidth]{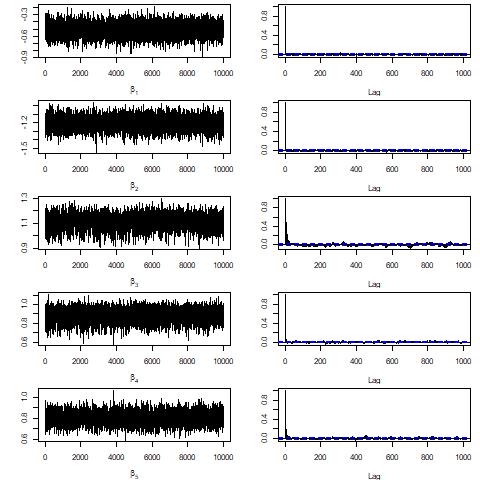}}
\caption{Trace plots and the sample ACF plots of $10^4$  MCMC iterations from posterior distributions using the algorithm described in subsection \ref{mcmc} for coefficient parameters of response variable $Y_2$.}\label{fig2}
\end{figure}

\begin{figure}[hbt!]
\centering{
\includegraphics[width=1\textwidth]{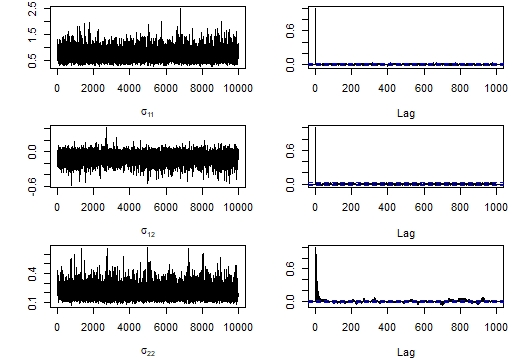}}
\caption{Trace plots and the sample ACF plots of $10^4$  MCMC iterations from posterior distributions using the algorithm described in subsection \ref{mcmc} for $\pmb{\Sigma}$.}\label{fig3}
\end{figure}

\end{document}